\documentclass[twocolumn,twocolappendix]{aastex63}

\usepackage{comment}
\usepackage{breqn}

\received{March 10, 2021}
\revised{April 25, 2021}
\accepted{\today}
\submitjournal{ApJ}

\shorttitle{Star formation rate function at $z\sim4.5$}
\shortauthors{Asada, Ohta \& Maeda}
\graphicspath{{./}{figures/}}

\begin{document}

\title{Star Formation Rate Function at $z\sim4.5$: An Analysis from rest UV to Optical}

\correspondingauthor{Yoshihisa Asada}
\email{asada@kusastro.kyoto-u.ac.jp}

\author[0000-0003-3983-5438]{Yoshihisa Asada}
\affiliation{Department of Astronomy, Kyoto University \\
Sakyo-ku, Kyoto 606-8502}

\author[0000-0003-3844-1517]{Kouji Ohta}
\affiliation{Department of Astronomy, Kyoto University \\
Sakyo-ku, Kyoto 606-8502}

\author[0000-0002-8868-1255]{Fumiya Maeda}
\affiliation{Department of Astronomy, Kyoto University \\
Sakyo-ku, Kyoto 606-8502}
\affiliation{Institute of Astronomy, Graduate School of Science, The University of Tokyo \\
Mitaka, Tokyo 181-0015}



\begin{abstract}

We present a star formation rate function (SFRF) at $z\sim4.5$ based on photometric data from rest UV to optical of galaxies in the CANDELS GOODS-South field using spectral energy distribution (SED) fitting.
We evaluate the incompleteness of our sample and correct for it to properly confront the SFRF in this study with those estimated based on other probes.
The SFRF is obtained down to $\sim10\ M_\odot\ \mathrm{yr}^{-1}$ and it shows a significant excess to that estimated from UV luminosity function and dust correction based on UV spectral slope.
As compared with the UV-based SFRF, the number density is larger by $\sim1$ dex at a fixed SFR, or the best-fit Schechter parameter of $\mathrm{SFR}^*$ is larger by $\sim1$ dex.
We extensively examine several assumptions on SED fitting to see the robustness of our result, and find that the excess still exist even if the assumptions change such as star formation histories, dust extinction laws, and one- or two-component model.
By integrating our SFRF to $0.22\ M_\odot\ \mathrm{yr}^{-1}$, the cosmic star formation rate density at this epoch is calculated to be $4.53^{+0.94}_{-0.87}\times10^{-2}\ M_\odot\ \mathrm{yr}^{-1}\ \mathrm{Mpc}^{-3}$, which is $\sim0.25$ dex larger than the previous measurement based on UV observations.
We also find that galaxies with intensive star formation ($>10\ M_\odot\ \mathrm{yr}^{-1}$) occupies most of the cosmic star formation rate density ($\sim80\%$), suggesting that star formation activity at this epoch is dominant by intensively star-forming galaxies.

\end{abstract}

\keywords{galaxies: evolution --- 
galaxies: formation --- galaxies: high-redshift}


\section{Introduction} \label{sec:intro}
Star formation rate function (SFRF) is one of the key properties of galaxies.
It directly describes the \textit{in-situ} evolution of galaxies at an epoch of the universe.
SFRF also provides the cosmic star formation rate density (CSFRD) which sheds light on the history of the universe.
Therefore, revealing SFRFs at various redshifts is crucial for the understanding of cosmological evolution of galaxies.

In high-$z$ ($z\gtrsim$4) universe, until recently, the SFRF and CSFRD are investigated mainly based on rest UV observations \citep[see e.g., a review by][]{madau_cosmic_2014}.
In estimating SFRF and CSFRD from UV luminosity function (LF), dust extinction is a major concern.
Since UV light emitted from massive stars can be easily attenuated by dust, it is important to correct for the loss of the light due to the dust extinction.
This correction is usually made by using a relation by \citet{meurer_dust_1999} which links a rest UV spectral slope $\beta$ and the amount of dust extinction.
Since FIR can probe dust-obscured star formation activity, FIR observations can also examine the CSFRD.
Some studies claim that the dust-obscured galaxies contribute to the CSFRD largely in high-$z$ ($z\gtrsim$4) universe
\citep[e.g.,][]{rowan-robinson_star_2016}, while others claim that the contribution is negligible
\citep[e.g.,][]{koprowski_evolving_2017}.
Thus the consensus on the evolution of CSFRD has not been reached yet, and
an independent estimation of SFRF/CSFRD is desirable.

Since the dust extinction has much less effect on rest optical light, utilizing not only rest UV but also optical data is expected to derive properties of galaxies more reliably.
Thus, deriving SFRF with data from rest UV to optical can be an independent estimation.
Recent deep observations by Infrared Array Camera (IRAC) on \textit{Spitzer} enable us to access the rest optical information of high-$z$ galaxies.
In determining properties of a galaxy with data from rest UV to optical, spectral energy distribution (SED) fitting is a common way.
Previous such studies using the rest UV to optical data also found an inconsistency between properties such as SFR or dust extinction derived from UV-based analysis and those from rest UV to optical data \citep[e.g.,][]{shim_z_2011,duncan_mass_2014,de_barros_properties_2014}.
The inconsistency leads to a systematic difference in SFRF.

However, the SFRF derived from rest UV to optical data is not investigated extensively or statistically evaluated.
In SED fitting, it is known that the dust reddening and red color due to aging of stellar population are degenerated, which makes the SFR derived from SED fitting less reliable.
Furthermore, if strong emission lines such as H$\alpha$ or [O\textsc{iii}]$\lambda5007$ from high-$z$ galaxies are redshifted into the IRAC band, it can boost IRAC broadband photometry \citep[e.g.,][]{yabe_stellar_2009,stark_keck_2013}.
The excess in the IRAC band may be interpreted as the presence of the dust reddening and/or old stellar population \citep[e.g.,][]{katz_probing_2019}.
This may also lead to the larger and/or smaller SFR.
These problems make it difficult to derive SFRF based on SED fitting and the previous studies did not take into account these aspects well enough.

In this study, we aim to derive SFRF based on the data from rest UV to optical considering these problems.
If an excess by emission line is seen in IRAC 4.5 $\mu$m band, it would be hard to recognize it as the emission line since the sensitivities of IRAC 5.8 $\mu$m and 8.0 $\mu$m bands are very much shallower than those in 3.6 $\mu$m and 4.5 $\mu$m bands.
To avoid this problem, targeting redshift around 4.5 and including nebular emissions in model spectrum are desirable.
The excess in 3.6 $\mu$m band due to a strong H$\alpha$ emission is expected to be recognized with 4.5 $\mu$m photometry.
Furthermore, we extensively examine assumptions on SED fitting that can have effects on the SFRF, such as various star formation histories (SFHs), two-component model (i.e., model composed of old stellar population and young star-forming population), and dust extinction law, to see the robustness of our result.
The incompleteness of our sample is corrected to derive SFRF, and the completeness limit of the SFRF is also evaluated.
These enable us to derive the SFRF from SED fitting which can be properly confronted with the SFRF estimated from UV LF.

This paper is structured as follows. In Section \ref{Sec:data}, we describe data and sample selection.
In Section \ref{Sec:SED_fitting}, details of the SED fitting are provided.
In Section \ref{Sec:res}, we present the resulting SFRF and see the effects by changing several assumptions.
Using the resulting SFRF, CSFRD is calculated in Section \ref{Sec:CSFRD}.
Section \ref{Sec:Summary} gives the summary of this paper.
Throughout this paper, all magnitudes are quoted in the AB system \citep{oke_secondary_1983}, and we assume the cosmological parameters of $H_0=70$ km s$^{-1}$ Mpc$^{-1}$, $\Omega_m = 0.3$ and $\Omega_\Lambda = 0.7$.

\section{data}\label{Sec:data}
\subsection{Photometric catalog and Photometric redshifts}
Among the surveys conducted with \textit{Spitzer}, Great Observatories Origins Deep Survey (GOODS) South field is one of the deepest and widest fields \citep[e.g.,][]{bradac_high-redshift_2020}.
We focus on this field and use the photometry catalog given by \citet{guo_candels_2013}, which is a UV to mid-infrared multiwavelength catalog in the Cosmic Assembly Near-infrared Deep Extragalactic Legacy Survey (CANDELS) GOODS South field.

The details about the object extraction and photometry are presented by \citet{guo_candels_2013}, so here we briefly summarize the method.
The catalog contains multi-wavelength band photometry consisting of observations by the Advanced Camera for Surveys (ACS) and Wide Field Camera 3 (WFC3) on \textit{Hubble Space Telescope} (\textit{HST}), IRAC on \textit{Spitzer}, and other ground-based observatories.
The optical data (ACS) contain observations in $F435W$, $F606W$, $F775W$, $F814W$ and $F850LP$ bands.
The NIR data (WFC3) contain observations in $F098M$, $F105W$, $F125W$ and $F160W$ bands.
From IRAC observations, 3.6 $\mu$m (Ch1), 4.5 $\mu$m (Ch2), 5.8$\mu$m (Ch3) and 8.0$\mu$m (Ch4) band photometries are available.
In addition, the catalog contains VIMOS and CTIO $U$-band data, and ISAAC and HAWK-I $K_s$-band data.

Photometry for \textit{HST} bands was conducted by using \textsc{SExtractor}'s dual-image mode.
Combined max-depth mosaic of $F160W$-band image ($H_{160}$ hereafter) was used for object detection and the photometry in other band was made with PSF matched image.
For the ground-based and \textit{Spitzer} images, photometry was done through \textsc{TFIT}.
In this work, we use all of the 17 band data but for CTIO $U$ band data from this photometry catalog, because the band was revealed to have red leak \citep[][]{guo_candels_2013}.

As for the redshifts, we utilize the CANDELS Bayesian photometric redshift catalog \citep{dahlen_critical_2013}.
For the entire objects in this catalog, photometric redshifts are derived based on a hierarchical Bayesian approach that combines the full $P(z)$ distributions derived by several manners.

\subsection{Sample Selection} \label{subsec:sample_sel}
\begin{figure}[tpb]
\plotone{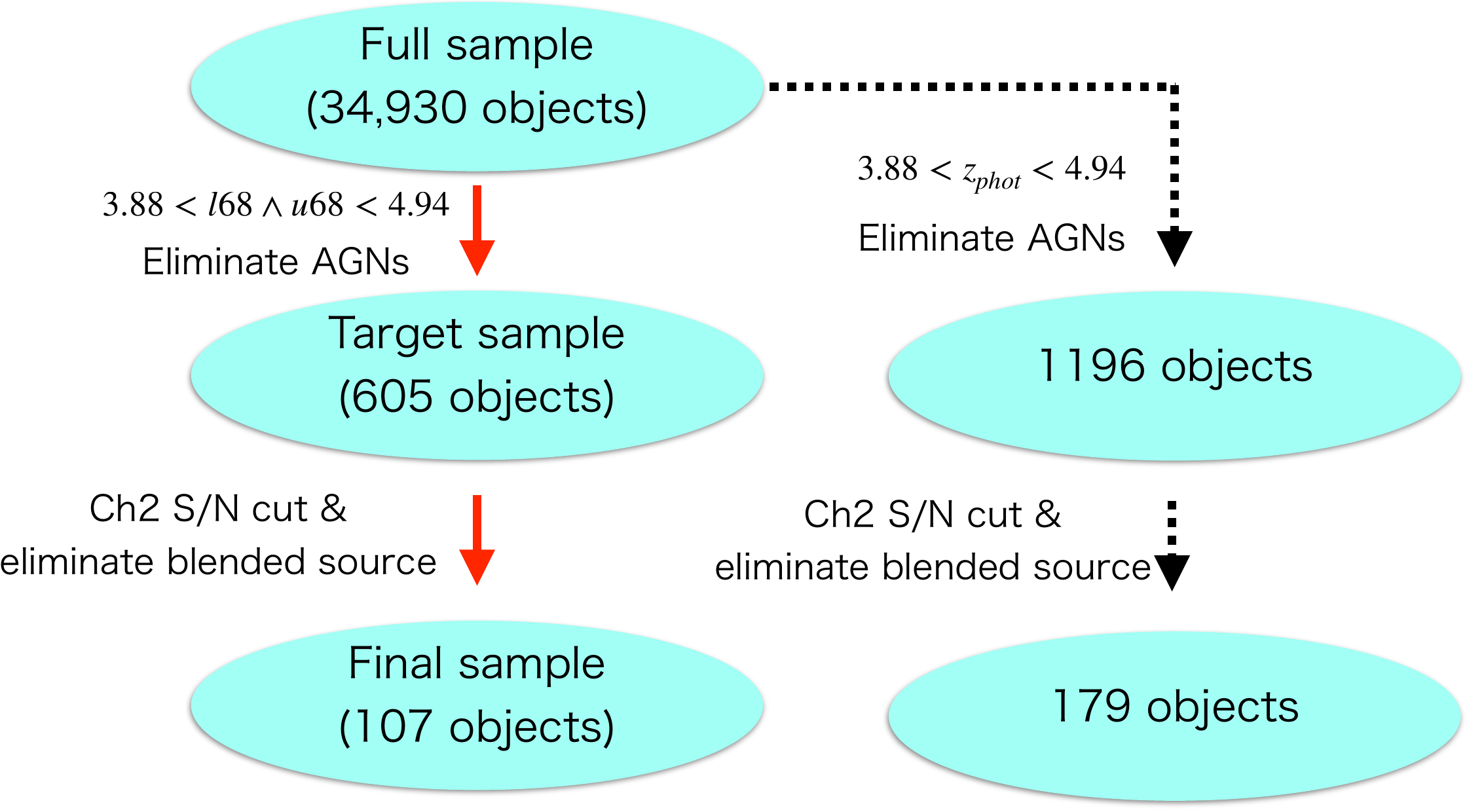}
\caption{Flow chart of our sampling. Red solid arrows indicate our selection, and the black dotted arrows indicate an alternative selection we use to test the effect of our criteria (Section \ref{Sec:CSFRD}).}
\label{sample_flow}
\end{figure}

In this study, to make a sample of galaxies at $3.88<z<4.94$ whose H$\alpha$ emission is redshifted into the \textit{Spitzer}/IRAC 3.6 $\mu$m band, we set the criteria composed of two steps.
Figure \ref{sample_flow} shows a flow chart of the criteria.
We adopt the route indicated by red solid arrows.

The CANDELS GOODS-S catalog contains 34,930 objects.
Among them, we extract objects whose redshift with $1\sigma$ confidence\footnote{The 1$\sigma$ width at this redshift is typically $\Delta z_{ph} \sim 0.27$} level is in the range of $3.88<z<4.94$:
we pick out objects that meet the following criteria,
\begin{equation} \label{photo-z_criteria}
    (3.88 < l68) \land (u68 < 4.94)
\end{equation}
where $l68$ and $u68$ is the lower and upper photo-$z$ 68\% confidence limit, respectively. 
Here, we exclude AGNs identified by \citet{hsu_candelsgoods-s_2014}.
The X-ray sensitivity limit of this AGN catalog is typically $3.2\times10^{-17}$, $9.1\times10^{-18}$ and $5.5\times10^{-17}$ erg s$^{-1}$ cm$^{-2}$ for full (0.5-8 keV), soft (0.5-2 keV) and hard (2-8 keV) band, respectively.
With the redshift of $z=4.5$, this limit corresponds to $\sim6.5\times10^{42}$, $\sim1.9\times10^{42}$ and $\sim1.1\times10^{43}$ erg s$^{-1}$, respectively.

605 objects pass the selection above (we refer to them as "target sample").
Next, we apply an additional cut to the target sample to ensure the SEDs of the galaxies in our sample are reliable to make SED fitting.
As we introduced in \S\ref{sec:intro}, detection in the 4.5 $\mu$m band of IRAC plays an important role in recognize the excess in the 3.6 $\mu$m band due to an H$\alpha$ emission.
Thus, we require a signal-to-noise ratio (S/N) larger than 5 in 4.5 $\mu$m band of IRAC.
In addition, we remove all the objects whose photometry can be contaminated by neighboring objects in IRAC images.
Specifically, we first discard objects whose separation from the nearest object is $<2^{\prime\prime}$ in $H_{160}$ image.
If the neighboring object is extended, photometry can be contaminated by the neighbor even if the separation from the nearest object is larger than $2^{\prime\prime}$.
Thus, for the objects whose separation in $H_{160}$ image is larger than $2^{\prime\prime}$, we conduct visual inspection on IRAC images whether the neighbors around the objects affect the photometry.

As a result, we make a sample containing 107 galaxies (we refer to this sample as "final sample").

\section{SED fitting}\label{Sec:SED_fitting}
We then perform SED fitting to the final sample.
To make SED model, we use population synthesis code \textsc{P\'egase}.3 \citep{fioc_pegase3_2019}.
This code includes nebular emission\footnote{We include the nebular emission from star forming clouds.} and follows the chemical evolution of the galaxy, which is then used to determine the metallicity of ISM\footnote{The metallicity of ISM is taken into account in determining the ratio of nebular emission lines.} and newly born stars at every time step.
The Chabrier03 IMF \citep{chabrier_galactic_2003} with the mass range of $0.08-120 M_\odot$ is adopted.
As for SFH, we adopt exponentially declining ($\propto e^{-t/\tau}$) and delayed exponential ($\propto t e^{-t/\tau}$) history with $\tau = 10,100,1000$ Myr, and constant star formation (CSF).
We also examine a two-component model later (Section \ref{subsubsec:twocomp}).
The universe at $z\sim4.5$ is aged $\sim1.4$ Gyr, we allow the age to vary from 1 Myr to 2 Gyr with 70 steps, which is not equally spaced in linear or logarithmic space.
Dust extinction is modeled with Calzetti law \citep{calzetti_dust_2000} with $R_V = 4.05$, but we will explore an alternative extinction law measured in the Small Magellanic cloud \citep[SMC;][]{pei_interstellar_1992} later (Section \ref{subsubsec:SMC}).
The color excess $E(B-V)$ is taken from 0.0 to 0.8 mag at an interval of 0.01 mag.
The ratio between stellar and nebular extinction is assumed to be $1$.
We adopt the default value of \textsc{P\'egase}.3 for the escape fraction.
Intergalactic attenuation by neutral hydrogen is applied following the prescription by \citet{madau_radiative_1995}.
The redshift of model galaxy is set from 3.8 to 5.0 at an interval of 0.1.
Consequently, we make 5,670 SED templates for each of the redshift steps and SFHs, and search for the best-fit SED for each galaxy by $\chi^2$ minimization.
Here, we do not fix the SFH but fix the redshift of the template to $z_{best}$ of the galaxy given by the CANDELS catalog.
Here, $z_{best}$ is basically the photometric redshift, but is spectroscopic redshift when it is available.
We do not use the photometry at the wavelengths shortward of Ly$\alpha$ in calculating the $\chi^2$. 

\begin{figure*}[tpb]
\plotone{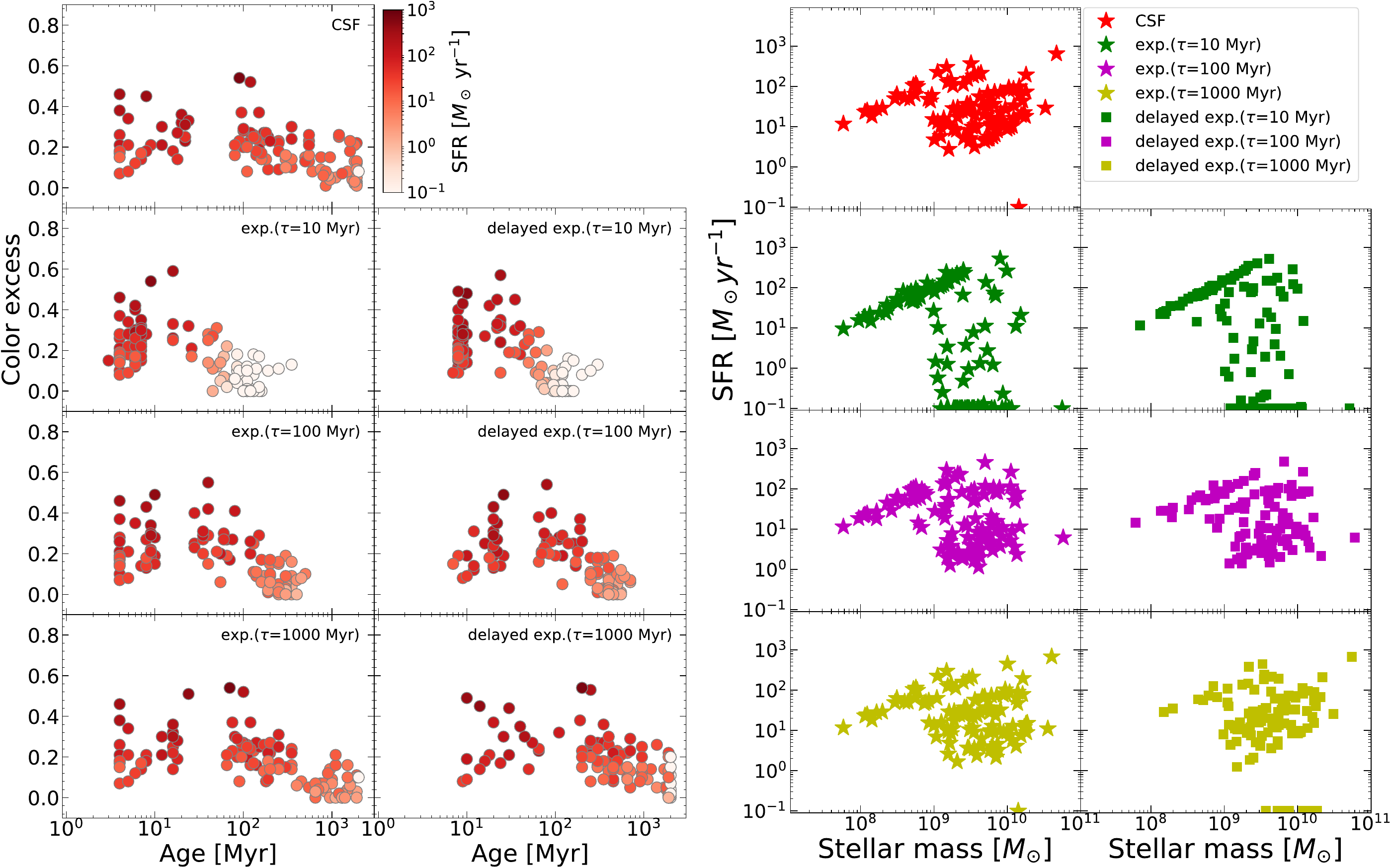}
\caption{Results of SED fitting.
\textit{Left}: In each panel, we show the distribution of best fit age and color excess $E(B-V)$ for all the objects in the final sample.
SFHs are different from panel to panel.
The color of the symbol refers to the value of SFR.
For clarity, SFRs smaller than $10^{-1}$ $M_\odot$ yr$^{-1}$ are treated as $10^{-1}$ $M_\odot$ yr$^{-1}$.
\textit{Right}: SFR vs. stellar mass is shown in each panel.
Red star symbol shows the result of CSF model. Green, magenta and yellow correspond to the value of $\tau=$10, 100, 1000 Myr respectively, and star and square shape correspond to exponentially declining and delayed exponential model, respectively.
}
\label{SED_fit_results}
\end{figure*}

Figure \ref{SED_fit_results} shows the resulting color excess $E(B-V)$ against age which is defined as the time since the onset of the star-formation.
The color excess tends to decrease with increasing age as expected.
However, the color excesses are not so large for ages of less than $10^{1.5}$ Myr.
This stems from the inclusion of the nebular continuum in our model spectra.
The color of the nebular continuum is redder than that of very young stellar continuum \citep[e.g.,][]{bouwens_very_2010}, thus the best-fit $E(B-V)$ tends to be smaller.
Figure \ref{SED_fit_results} (right) shows SFR against stellar mass.
The SFR is defined as the instantaneous value of the best-fit template.
The distribution is basically similar to that at $z\sim4.5$ derived by using SED fitting in the previous studies \citep[e.g.,][]{caputi_star_2017,faisst_recent_2019}.

\section{Star formation rate function}\label{Sec:res}
In this section, we first describe the method for deriving SFRF using the result of SED fitting (\S \ref{subsec:prop}). 
Next, we derive the SFRF and compare it with that estimated from UV LF (\S \ref{subsec:SFRF}).
In \S \ref{subsec:reason}, the reason for the difference between our SFRF and UV-based one is examined.
We also derive SFRF with various assumptions to see the effect by the difference of model assumption (\S \ref{subsec:otherSFRF}).
Finally, in \S \ref{subsec:further_insp}, we present several further inspections related to the result derived in \S \ref{subsec:SFRF}-\ref{subsec:otherSFRF}.

\subsection{Method for deriving SFRF} \label{subsec:prop}

In order to derive SFRF, incompleteness of the sample should be corrected properly.
The final sample is affected by 3 factors: (i) detection rate in $H_{160}$-band image, (ii) S/N cut in 4.5 $\mu$m band and (iii) elimination of blended objects with neighbors.

\begin{figure}[tpb]
\plotone{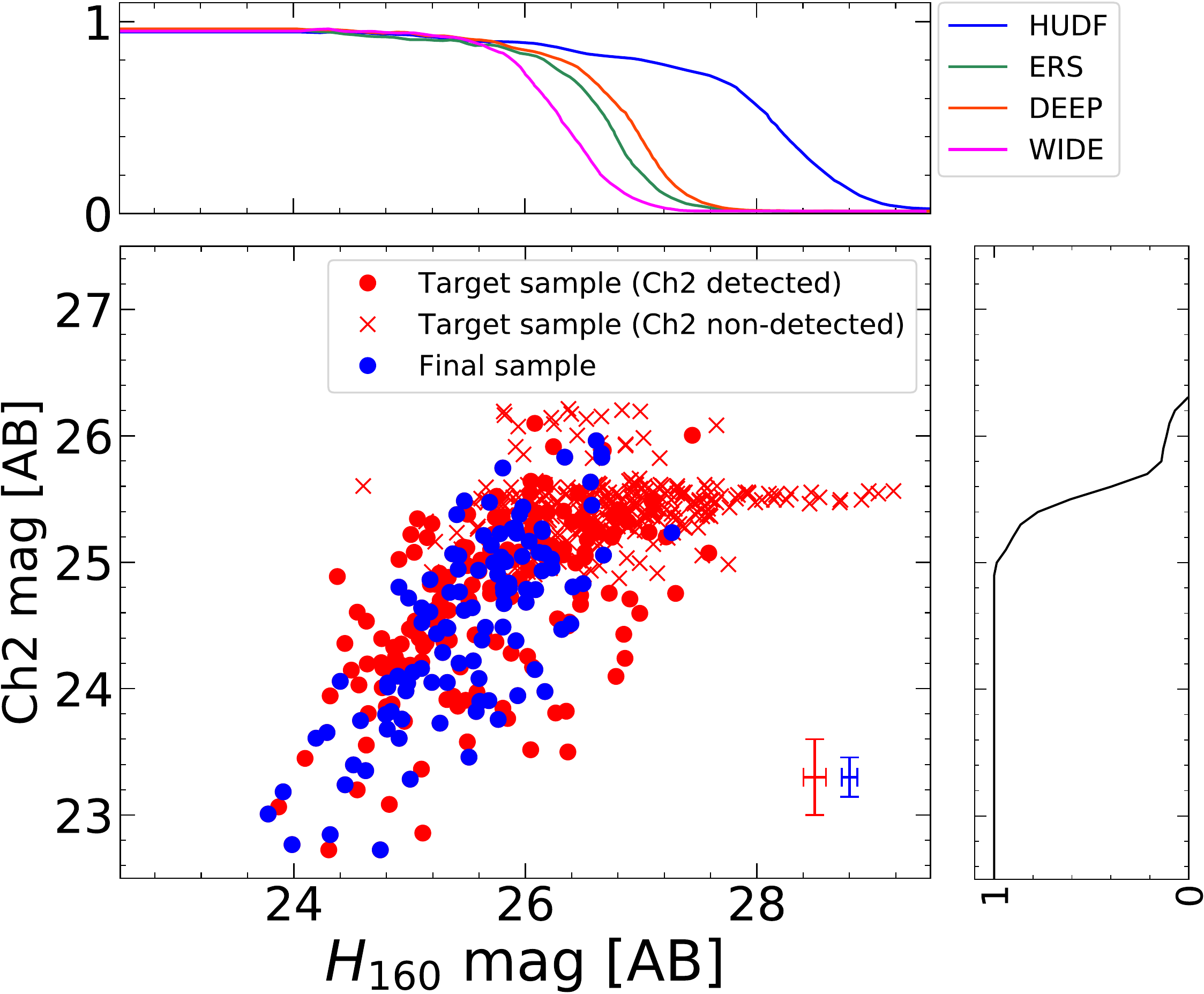}
\caption{\textit{Top}:Detection rate at $H_{160}$ band given by \citet{duncan_mass_2014}. Blue, red, green and magenta line correspond to HUDF, ERS, DEEP and WIDE, respectively. This function is given only for the magnitude fainter than $\sim24$ mag, so we extrapolate it brightward.
\textit{Lower left}: Apparent magnitudes at $H_{160}$ and 4.5 $\mu$m band distribution. 
Red points and crosses represent the galaxies in our target sample that are detected and not detected in 4.5 $\mu$m band, respectively.
Blue points represent the galaxies in our final sample.
The median of photometric uncertainty is shown in the right bottom.
\textit{Lower right}: Cumulative histogram of 5$\sigma$ limit magnitude at 4.5 $\mu$m band as a function of 4.5 $\mu$m magnitude.}
\label{mag_dist}
\end{figure}

To correct for (i), the detection rate in $H_{160}$ is required.
\citet{duncan_mass_2014} derived the detection rate of the CANDELS GOODS-S catalog by conducting mock observation, so we use their result.
Figure \ref{mag_dist} (top panel) shows the detection rate against apparent magnitudes at $H_{160}$.
Note that \citet{duncan_mass_2014} estimate the rate dividing the GOODS-S fields into 4 subregions according to the exposure time; HUDF, ERS, DEEP and WIDE.

We evaluate the detection rate in the IRAC 4.5 $\mu$m band to correct for (ii).
The exposure time in the surveyed region is inhomogeneous.
Thus we evaluate this effect as follows.
In the CANDELS photometric catalog, $1\sigma$ limiting magnitudes at the position of all the objects in each band are also available.
We calculate the $5\sigma$ limit at the location of the objects in our target sample and make a cumulative histogram of it.
This can be used for the correction for the difference of the limiting magnitudes since $z\sim4.5$ galaxies are almost randomly distribute in the whole survey region.
Figure \ref{mag_dist} (right panel) shows the distribution of 5$\sigma$ limit against apparent magnitudes at 4.5 $\mu$m band.

As for (iii), blending occurs regardless of its SFR as far as their apparent size is similar, so we can correct for it by dividing the fraction of isolated objects.
Therefore, we derive the fraction of isolated object in the IRAC image $f^{\mathrm{sel}}$ at $z\sim4.5$.
We first randomly pick out $\sim100$ objects from our target sample and see their images in the IRAC 4.5 $\mu$m band.
We categorize the $\sim100$ objects into 3 groups; objects clearly detected and not blended with the neighboring objects (group A), objects clearly detected but heavily contaminated by the neighbors (group B)\footnote{Objects whose separation from the neighbor is less than 2$^{\prime\prime}$ are included here.}, and objects that is faint and not detected.
We then calculate the fraction $f^{\mathrm{sel}}$ by $\#(A)/(\#(A)+\#(B))$, and obtain the value of $f^{\mathrm{sel}} \sim 0.35$.

By considering the effects of incompleteness due to these factors, the SFRF $\phi(\dot{M}_\star)$ [Mpc$^{-3}$ dex$^{-1}$] for the bin of $d(\log \dot{M}_\star)$ can be estimated as follows:
\begin{equation}\label{estimate_density_per_dex}
    \phi(\dot{M}_\star)d(\log \dot{M}_\star) = \sum^{N_{\mathrm{gal}}}_i \frac{1}{f^{\mathrm{sel}} f^{\mathrm{det}}(m_{ch2,i}) V_i^{\mathrm{eff}}}
\end{equation}
where the subscript $i$ represents each of the galaxies in the bin, $N_{\mathrm{gal}}$ is the number of them, $f^{\mathrm{sel}}$ is the fraction of isolated objects in the IRAC image (iii), $f^{\mathrm{det}}(m_{ch2,i})$ is the value in the cumulative histogram of 5$\sigma$ limit in 4.5 $\mu$m band at the magnitude of $m_{ch2,i}$ (ii) and $V^{\mathrm{eff}}_i$ is the effective volume of the survey for this galaxy $i$ (i).
This effective volume for a galaxy $i$ can be calculated as
\begin{equation}
    V^{\mathrm{eff}}_i = \sum^{N_{\mathrm{region}}}_k f^{160}_k(m_{160,i}) \Omega_k \int^{r_{z=4.94}}_{r_{z=3.88}} r^2 dr
\end{equation}
where the summation, $k$, is over the subregions in the field, $N_{\mathrm{regions}}$ is the number of them ($=4$), $f^{160}_k(m_{160,i})$ is the detection rate in $H_{160}$ band at the magnitude of $m_{160,i}$, $\Omega_k$ is the solid angle of the subregion, and $r_{z=a}$ is the comoving distance from $z=0$ to $z=a$.

\begin{figure}[tpb]
\includegraphics[width=1.0\columnwidth, angle=0]{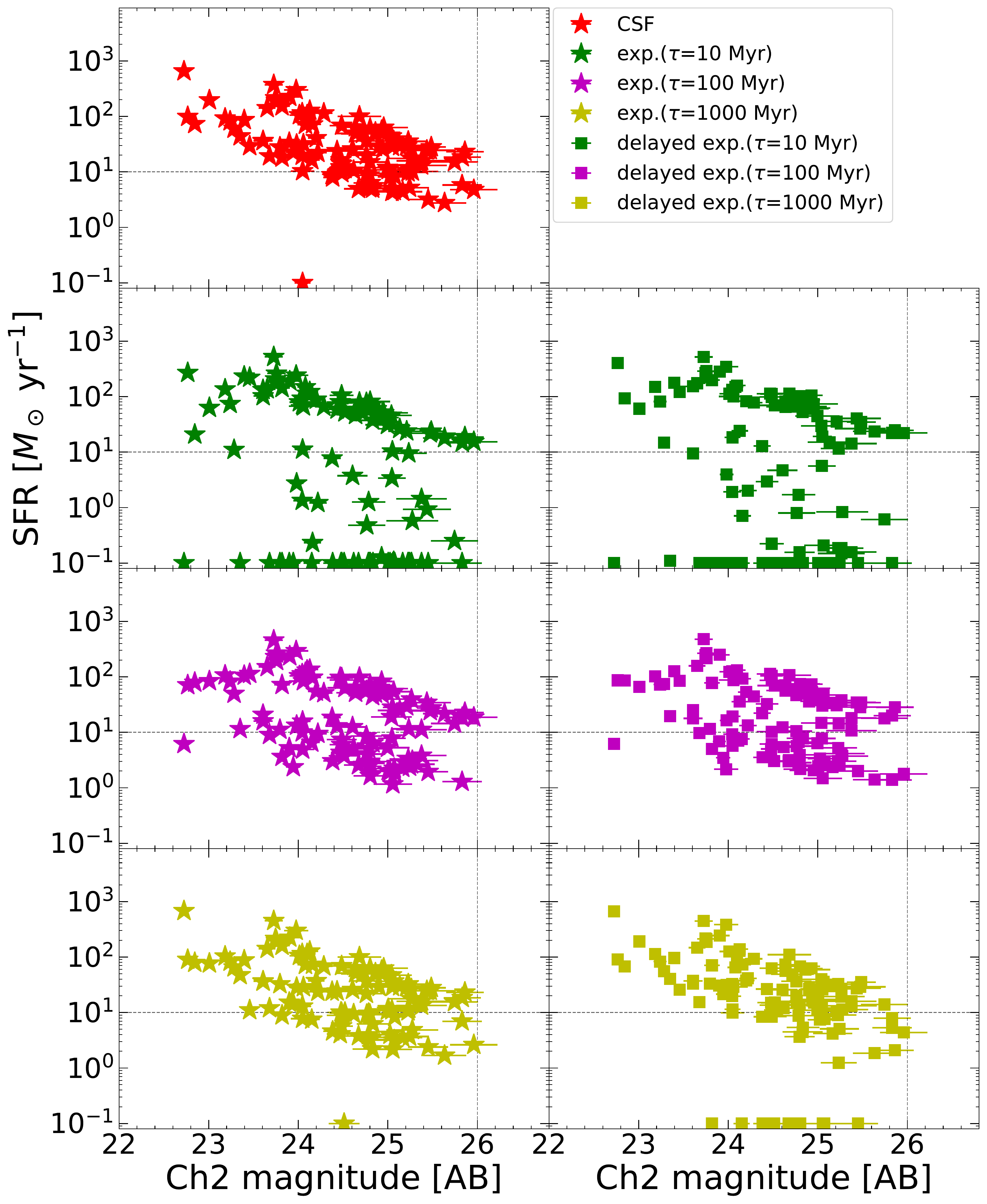}
\caption{SFR vs. 4.5 $\mu$m magnitude.
SFH is different from panel to panel. Red star symbol shows the result of CSF model. Green, magenta and yellow correspond to the value of $\tau=$10, 100, 1000 Myr respectively, and star and square shape correspond to exponentially declining and delayed exponential model, respectively (other than red star symbol).
We also show guide lines that correspond to $m_{ch2}=26.0$ mag and $\mathrm{SFR}=10\ M_\odot\ \mathrm{yr}^{-1}$ (see text for the detail).
For clarity, SFRs smaller than $10^{-1}$ $M_\odot$ yr$^{-1}$ are plotted as $10^{-1}$ $M_\odot$ yr$^{-1}$.}
\label{SFRvsCh2}
\end{figure}

Additionally, we intend to estimate the completeness limit of SFR.
We apply S/N cut in 4.5 $\mu$m band, and the 4.5 $\mu$m magnitude of a galaxy at about this redshift is a good indicator of its stellar mass \citep[e.g.,][]{yabe_stellar_2009}.
Thus, our sample is expected to be stellar mass limited.
Since a rough correlation is seen between the SFR and stellar mass (Figure \ref{SED_fit_results} right), this limit would also be SFR limit.
To see this, we examine the correlation between SFR and 4.5 $\mu$m magnitude.
Figure \ref{SFRvsCh2} shows the SFR versus 4.5 $\mu$m magnitude.
A trend that a brighter galaxy shows a larger SFR for SF galaxies is seen regardless of SFH.
Since our detection limit in 4.5 $\mu$m band is $\sim 26.0$ mag (lower right panel in Figure \ref{mag_dist}) and this magnitude corresponds to SFR $\sim10$ $M_\odot$ yr$^{-1}$ (Figure \ref{SFRvsCh2}), the SFRF is constructed down to SFR $\sim10$ $M_\odot$ yr$^{-1}$ by correcting for the incompleteness discussed above.
One might worry about the effect by object that is bright enough to be detected in 4.5 $\mu$m band, which in turn its SFR can be larger than the SFR completeness limit $10$ $M_\odot$ yr$^{-1}$, but too faint in $H_{160}$ band.
However, considering the distribution of the magnitudes in $H_{160}$ and 4.5 $\mu$m band (lower left panel in Figure \ref{mag_dist}), such objects should be rare and the effect on SFRF is expected to be negligible.

\subsection{Star Formation Rate Function}\label{subsec:SFRF}

\begin{figure}[tpb]
\includegraphics[width=1.0\columnwidth, angle=0]{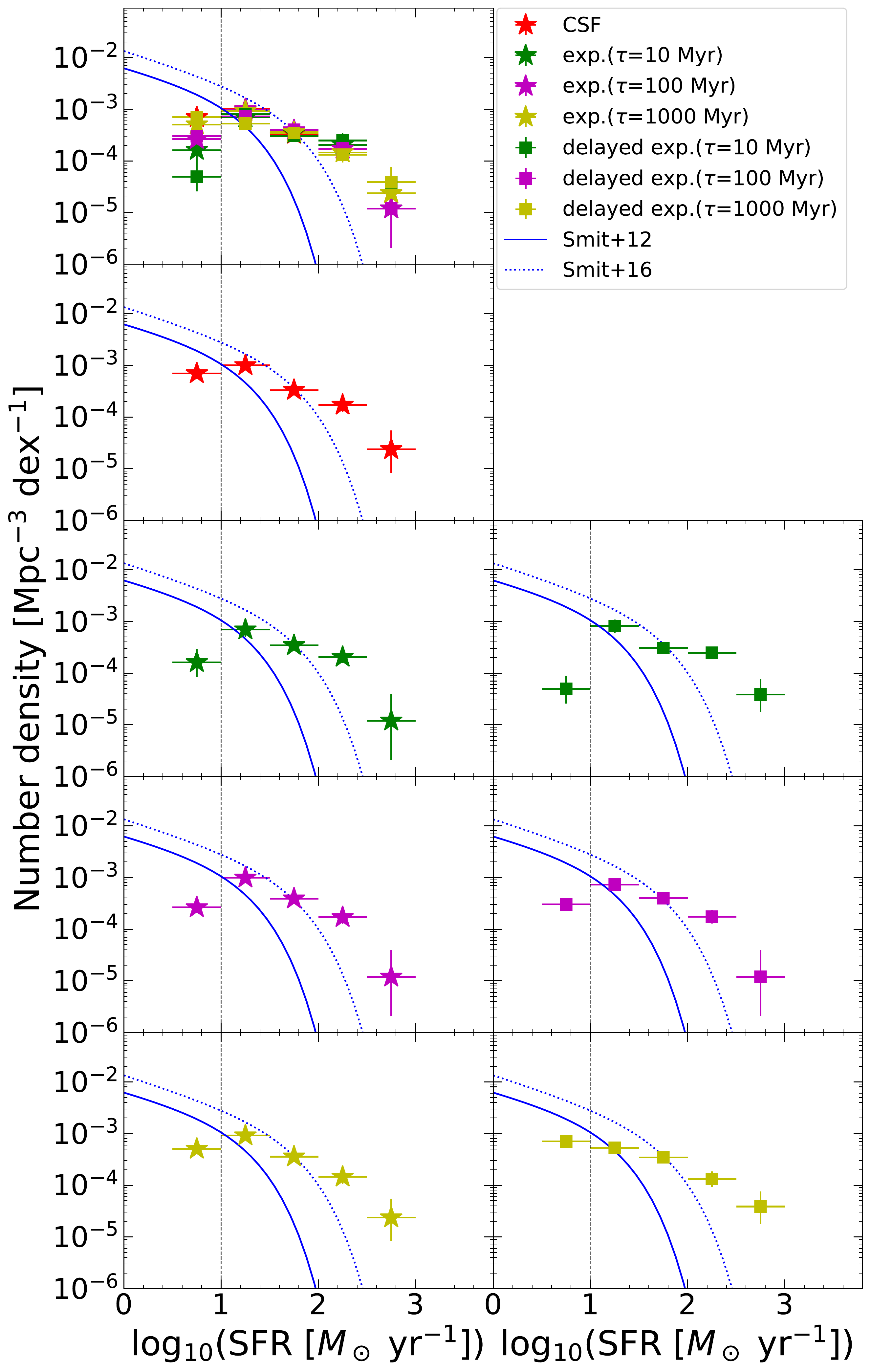}
\caption{SFRF at $z\sim4.5$. Panels and symbols are the same as Figure \ref{SFRvsCh2}.
The most upper panel shows the results from all the SFHs.
Blue solid and dotted lines are SFRF at $z\sim5$ given by \citet{smit_star_2012} (S12) and \citet{smit_inferred_2016} (S16), respectively, which are derived from UV LFs after correcting for the dust extinction.
Black dotted line is a guide line corresponding to $\mathrm{SFR}=10\ M_\odot\ \mathrm{yr}^{-1}$.
The vertical error bars are $1\sigma$ Poisson uncertainties by \citet{gehrels_confidence_1986}.}
\label{SFRF_indev}
\end{figure}

\begin{figure*}[tpb]
\plotone{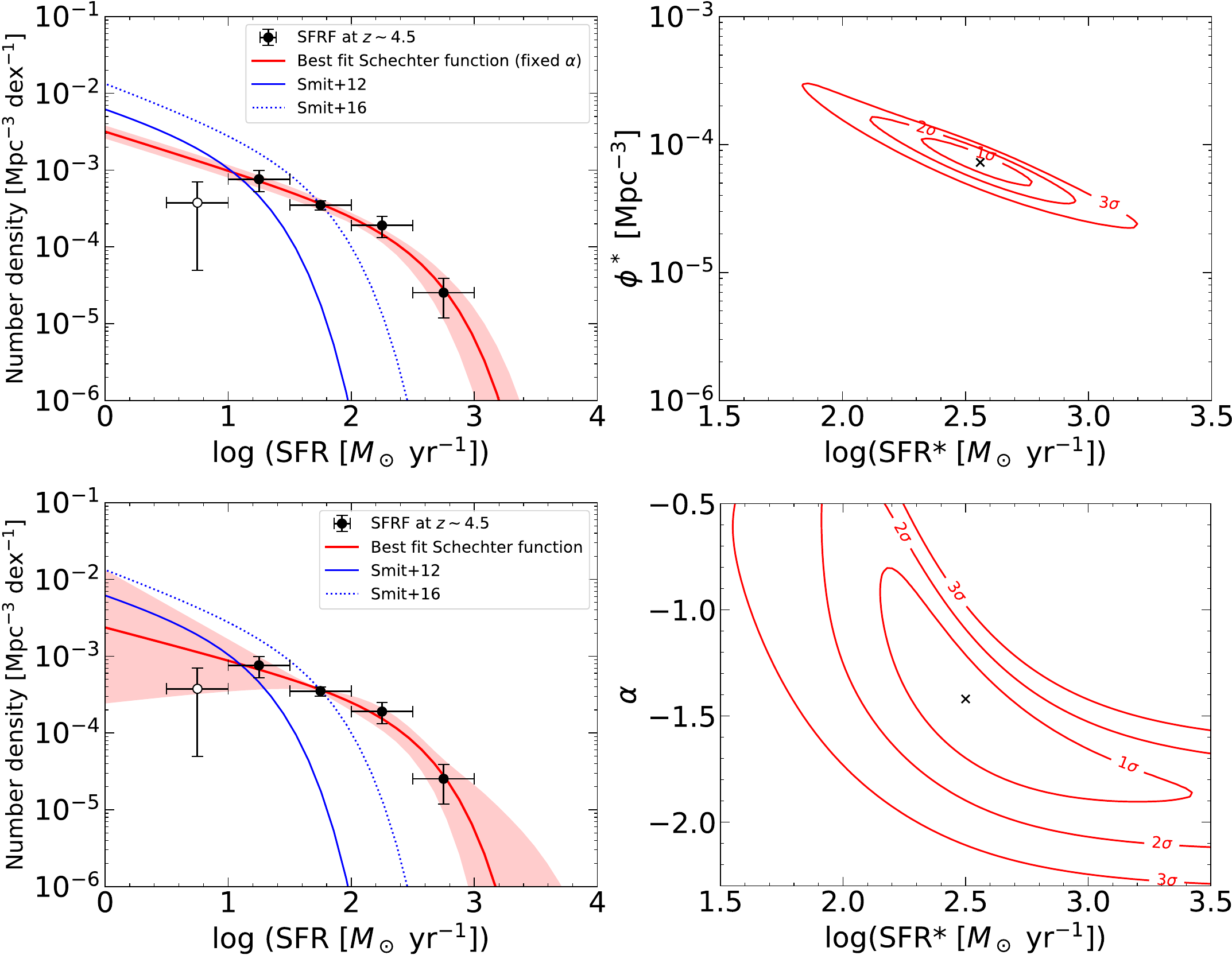}
\caption{SFRF obtained from the SED fitting (black circles) and the best-fit Schechter function (red solid line).
The black filled and open circle represent the data points that are used and not used in the fitting, respectively.
\textit{Upper left}: Fixed $\alpha$ with red shaded region representing 1$\sigma$ error.
\textit{Upper right}: Error contours for Schechter parameters $\phi^*$ and $\log_{10}(\mathrm{SFR}^*)$. Black cross represents the best-fit values.
\textit{Lower left}: Same as the upper left panel but for fitting without fixing $\alpha$.
\textit{Lower right}: Same as the upper right panel but for fitting without fixing $\alpha$. Note that this contour in the lower right panel is for the case where $\phi^*$ is fixed to its best-fit value.
Blue solid and dotted lines show the SFRFs converted from UV LF at $z\sim5$ (see text for details).
}
\label{SFRF_fitting}
\end{figure*}

The resulting SFRF is shown in Figure \ref{SFRF_indev}.
As we can see in Figure \ref{SFRF_indev}, the SFRFs with various SFHs agree well with each other in the complete region ($>10$ $M_\odot$ yr$^{-1}$).
Given their error bars are smaller than the differences among SFH models,
we adopt the maximum and minimum value in each bin among these SFHs as the upper and lower limit of the uncertainty for SFRF, respectively, and adopt the average values of these maximum and minimum as the fiducial values of SFRF.
The fiducial SFRF is shown in Figure \ref{SFRF_fitting}.

For comparison, we plot the SFRF at $z\sim5$ by \citet{smit_star_2012} (S12, hereafter) converted to the same IMF as we use.
This SFRF by S12 is constructed by correcting for observed UV luminosity function with the \citet{meurer_dust_1999} IRX-$\beta$ relation.
We also plot the SFRF presented by \citet{smit_inferred_2016} (S16, hereafter).
S16 found a systematic offset between H$\alpha$- and UV-based SFR at $z\sim4.5$.
To resolve this tension in the inferred SFRs, they examined the impact of the assumed dust extinction and stellar properties on the inferred SFRs.
They consequently proposed two types of SFRFs: \citet{meurer_dust_1999} type and SMC type.
Since \citet{meurer_dust_1999} dust correction is almost identical to that with Calzetti law, we plot \citet{meurer_dust_1999} type model for fair comparison.

The SFRF obtained in this study obviously shows a large excess to the SFRF estimated from UV LF by S12 and S16.
The SFRF from UV LF by S16 is made to reconcile the discrepancy between SFRs derived from rest UV and H$\alpha$ at this redshift, though, it still seems to underestimate the number of galaxies with large SFR.
We will deal with this difference between our result and rest UV-based SFRF in the next subsection.

When we compare the resulting SFRF with that estimated from FIR observation, the best-fit SFRF at $z=4.25$ by \citet{rowan-robinson_star_2016} delineates the upper envelope of our SFRF.
Although the SFRF by \citet{rowan-robinson_star_2016} is derived only in the largest SFR region ($\log_{10}(\mathrm{SFR}\ [M_\odot\ \mathrm{yr}^{-1}])>3.5$) and their best-fit SFRF is extrapolated down to $\log_{10}(\mathrm{SFR}\ [M_\odot\ \mathrm{yr}^{-1}])\sim1.5$ by assuming a \citet{saunders_60-micron_1990} functional form, our result is roughly consistent with it.

We fit the Schechter function to our data for SFRF through $\chi^2$ minimization.
Schechter function can be written as
\begin{dmath}
    \phi(\dot{M}_\star) d(\log\dot{M}_\star)= (\ln10)\phi^* \left(\frac{\dot{M}_\star}{\mathrm{SFR}^*}\right)^{\alpha+1}\exp\left[-\frac{\dot{M}_\star}{\mathrm{SFR}^*}\right] d(\log\dot{M}_\star)
\end{dmath}
where $\mathrm{SFR}^*,\alpha,\phi^*$ is the characteristic SFR, faint-end slope and characteristic number density, respectively.
Because our SFRF is complete only in the large SFR region, it is difficult to determine the parameter $\alpha$.
Thus, we tentatively fix the value as $\alpha=-1.50$, which is the same as that of SFRF at $z\sim5$ estimated from UV LF (S12).
We also conduct fitting leaving $\alpha$ as a free parameter for comparison.
We only use data points at SFR $>10$ $M_\odot$ yr$^{-1}$ in the fitting.
When the parameter $\alpha$ is fixed to be $-1.50$, the best-fit Schechter parameters are $\phi^* = 7.24^{+1.07}_{-1.36}\times10^{-5}$ Mpc$^{-3}$ and $\log_{10}(\mathrm{SFR}^* [M_\odot$ yr$^{-1}]) = 2.56^{+0.10}_{-0.12}$.
If we allow $\alpha$ to vary, the best-fit values are $\alpha=-1.42^{+0.24}_{-0.34}$, $\phi^* = 9.20^{+1.76}_{-1.61}\times10^{-5}$ Mpc$^{-3}$, and $\log_{10}(\mathrm{SFR}^* [M_\odot$ yr$^{-1}]) = 2.50^{+0.22}_{-0.28}$.
The fixed value of $\alpha$ is within its uncertainties.
These results are shown in Figure \ref{SFRF_fitting} and Table \ref{Tab1}.

\begin{deluxetable*}{ccccccc}
\tablenum{1}\label{Tab1}
\tablecaption{Best-fit Schechter parameters and Cosmic star formation rate density}
\tablewidth{0pt}
\tablehead{
\colhead{Redshift} & \colhead{$\alpha$} & \colhead{$\log_{10}$(SFR$^*$)} & \colhead{$\phi^*$} & \colhead{$\log_{10}\rho_{\mathrm{SFR}}$} & \colhead{Note} & \colhead{Dust extinction} \\
\colhead{} & \colhead{} & \colhead{($M_\odot$ yr$^{-1}$)} & \colhead{($10^{-5}$ Mpc$^{-3}$)} & \colhead{($M_\odot$ yr$^{-1}$ Mpc$^{-3}$)} & \colhead{} & \colhead{}
}
\decimalcolnumbers
\startdata
4.5 & $-1.50$ (fixed) & $2.56^{+0.10}_{-0.12}$ & $7.24^{+1.07}_{-1.36}$ & $-1.34^{+0.08}_{-0.09}$ & & Calzetti \\
4.5 & $-1.42^{+0.24}_{-0.34}$ & $2.50^{+0.22}_{-0.28}$ & $9.20^{+1.76}_{-1.61}$ & $-1.36^{+0.19}_{-0.10}$ & $\alpha$ is varied & Calzetti \\
4.5 & $-1.50$ (fixed) & $2.40^{+0.06}_{-0.10}$ & $11.0^{+1.6}_{-2.0}$ & $-1.33^{+0.07}_{-0.09}$ & Two comp. model & Calzetti \\
\hline
4.5 & $-1.56$ (fixed) & $2.22^{+0.16}_{-0.22}$ & $3.89^{+2.00}_{-1.80}$ & $-1.91^{+0.13}_{-0.16}$ & & SMC \\
4.5 & $-1.56$ (fixed) & $2.10^{+0.08}_{-0.10}$ & $13.5^{+3.1}_{-2.5}$ & $-1.50^{+0.10}_{-0.09}$ & Two comp. model & SMC
\enddata
\tablecomments{The star formation rate density is calculated by integrating SFRF down to $\sim 0.22$ $M_\odot$ yr$^{-1}$, which corresponds to $M_{\mathrm{UV}} = -17$ mag.}
\end{deluxetable*}

\subsection{What makes the difference of SFRF?}\label{subsec:reason}
In the previous subsection, we have demonstrated that SFRF from SED fitting shows a large excess to the SFRF from UV LF.
The main reason for this is considered to be the difference of estimated dust extinction.
In converting observed UV LF to SFRF, the dust correction is made by using \citet{meurer_dust_1999} IRX-$\beta$ relation:
\begin{equation}\label{IRX_M99}
    A_{1600} = 4.43 + 1.99\beta
\end{equation}
This relation links the observed spectral slope $\beta$ in rest UV and the rest UV extinction $A_{1600}$.
We measure the slope $\beta$ for all the objects in our final sample.
For the measurement of $\beta$, we simply conduct power-law fit to the photometry from $F814W$ band ($i_{814}$ hereafter) to $H_{160}$.
We compare the attenuation derived from SED fitting and this relation.
\begin{figure}[tpb]
\includegraphics[width=1.0\columnwidth, angle=0]{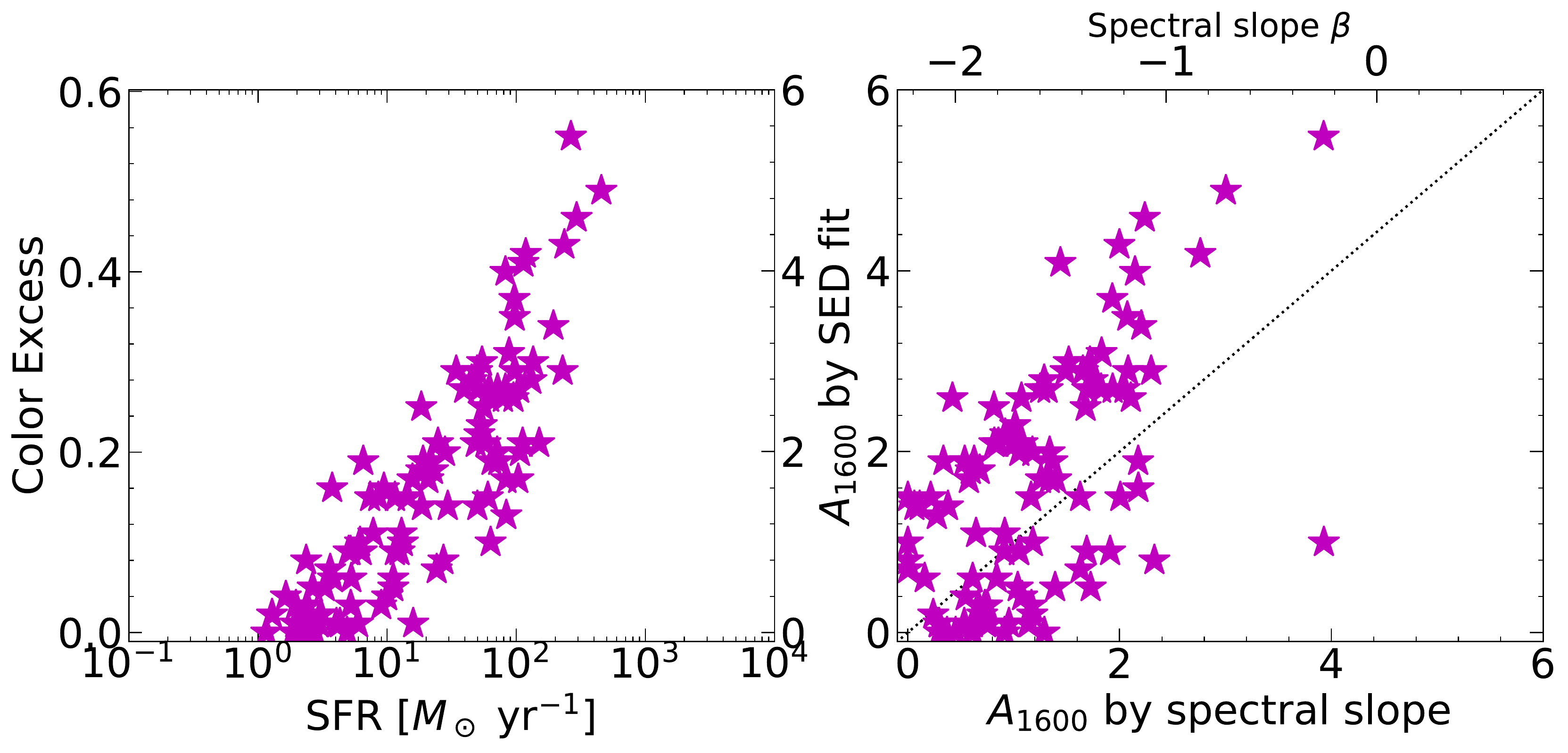}
\caption{\textit{Left}: Color excess ($E(B-V)$) against SFR derived from SED fitting.
Corresponding $A_{1600}$ is also shown at the right ordinate.
\textit{Right}: $A_{1600}$ from SED fitting and that from spectral slope $\beta$.
We use the equation (\ref{IRX_M99}) to convert spectral slope $\beta$ into the extinction $A_{1600}$.
Only the case of exponentially declining SFH model with $\tau=100$ Myr is shown here, but the distribution does not change so much with the choice of SFHs.
}
\label{A1600comparison}
\end{figure}

The results are shown in Figure \ref{A1600comparison}.
Galaxies with a larger SFR tend to show a larger color excess (left panel), and the dust extinctions of such objects are especially underestimated when we use IRX-$\beta$ relation (right panel).
We infer that, for a red galaxy, SED fitting gives larger value of $A_{1600}$ than that derived from the IRX-$\beta$ relation, and hence the intrinsic rest UV luminosity and SFR gets larger.

\begin{figure}[tpb]
\plotone{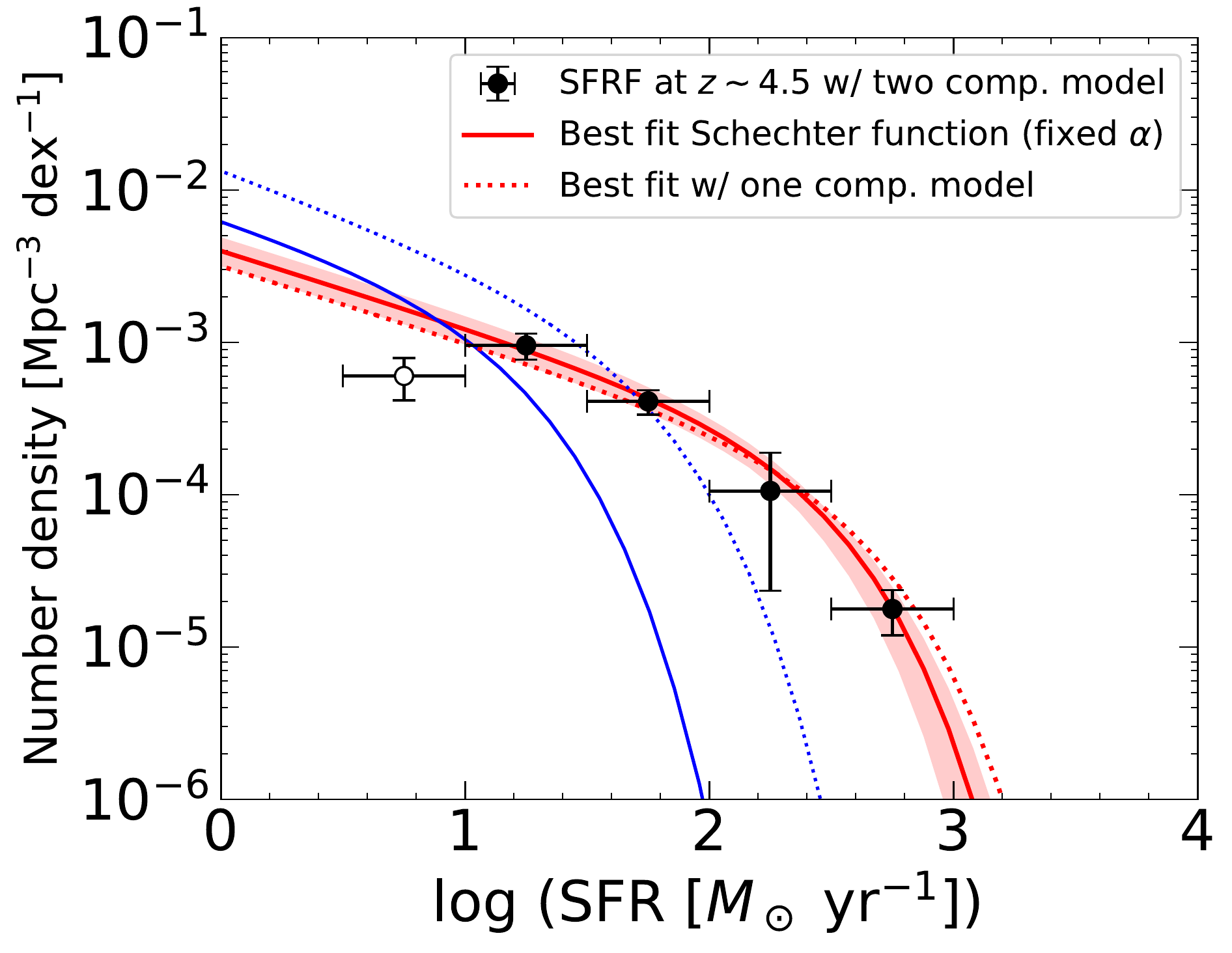}
\caption{Same as Figure \ref{SFRF_fitting} but for the two-component model and the best-fit Schechter function (fixed $\alpha$, red solid line).
Red dotted line shows the best-fit Schechter function to the SFRF with one-component model fixing $\alpha$ (c.f. upper left panel in Figure \ref{SFRF_fitting}).
Blue solid and dotted lines are the same as those in Figure \ref{SFRF_fitting}.}
\label{SFRF_fitting_twocomp}
\end{figure}

\subsection{SFRF with other assumptions}\label{subsec:otherSFRF}
In this subsection, we derive the SFRF by changing the assumptions on SED fitting to see the robustness of our result.
\subsubsection{Two-component model}\label{subsubsec:twocomp}
As we described in \S \ref{subsec:reason}, the excess of SFRF is mainly due to the difference of the estimation of dust extinction.
However, red color of SED can be explained not only by dust extinction but also by ageing of stellar population.
Therefore, we examine a two-component model composed of old stellar population and young star-forming population.
If the red color of the high-SFR objects can be explained not only by dust but also by old stellar component to some extent, the best-fit value of $E(B-V)$ and the discrepancy of SFRFs are expected to be reduced.
In order to examine this, we conduct SED fitting using the two-component model.
As for the old stellar population, we use a spectrum of 1 Gyr old quenched galaxy.
Details of this two-component analysis are presented in Appendix \ref{Appendix:Twocomp}.
Resulting SFRF is shown in Figure \ref{SFRF_fitting_twocomp} and Table \ref{Tab1}.
As seen in Figure \ref{SFRF_fitting_twocomp} and Table \ref{Tab1}, adopting two-component model can reduce the discrepancy only slightly, and the excess in large SFR region still exists even with the two-component model.

Interestingly, adopting this model not only reduces the number density at most intensive region (SFR $\gtrsim 10^2$ $M_\odot$ yr$^{-1}$) but also increases at intermediate region (SFR $\sim10^{1-2}$ $M_\odot$ yr$^{-1}$) (c.f. Appendix \ref{Appendix:Twocomp}), though the difference is not large.

\subsubsection{Alternative reddening law}\label{subsubsec:SMC}
The assumption of dust extinction curve can have systematic effect on the estimated SFR.
Thus, it is important to see the impact of the assumption of the extinction law on the SFRF.
So far, we adopt the Calzetti law for dust extinction.
However, some studies suggest that high-$z$ galaxies prefer SMC-like extinction curve \citep[e.g.,][]{fudamoto_alpine-alma_2020}.
Adopting SMC extinction law systematically reduces the value of SFR derived from SED fitting compared to that with Calzetti attenuation curve \citep[e.g.,][]{yabe_stellar_2009}.
We conduct SED fitting with the same procedure as we did in Section \ref{Sec:SED_fitting} assuming SMC extinction curve by \citet{pei_interstellar_1992}, and derive SFRF in the same way as Section \ref{subsec:SFRF}.

\begin{figure}[tpb]
\plotone{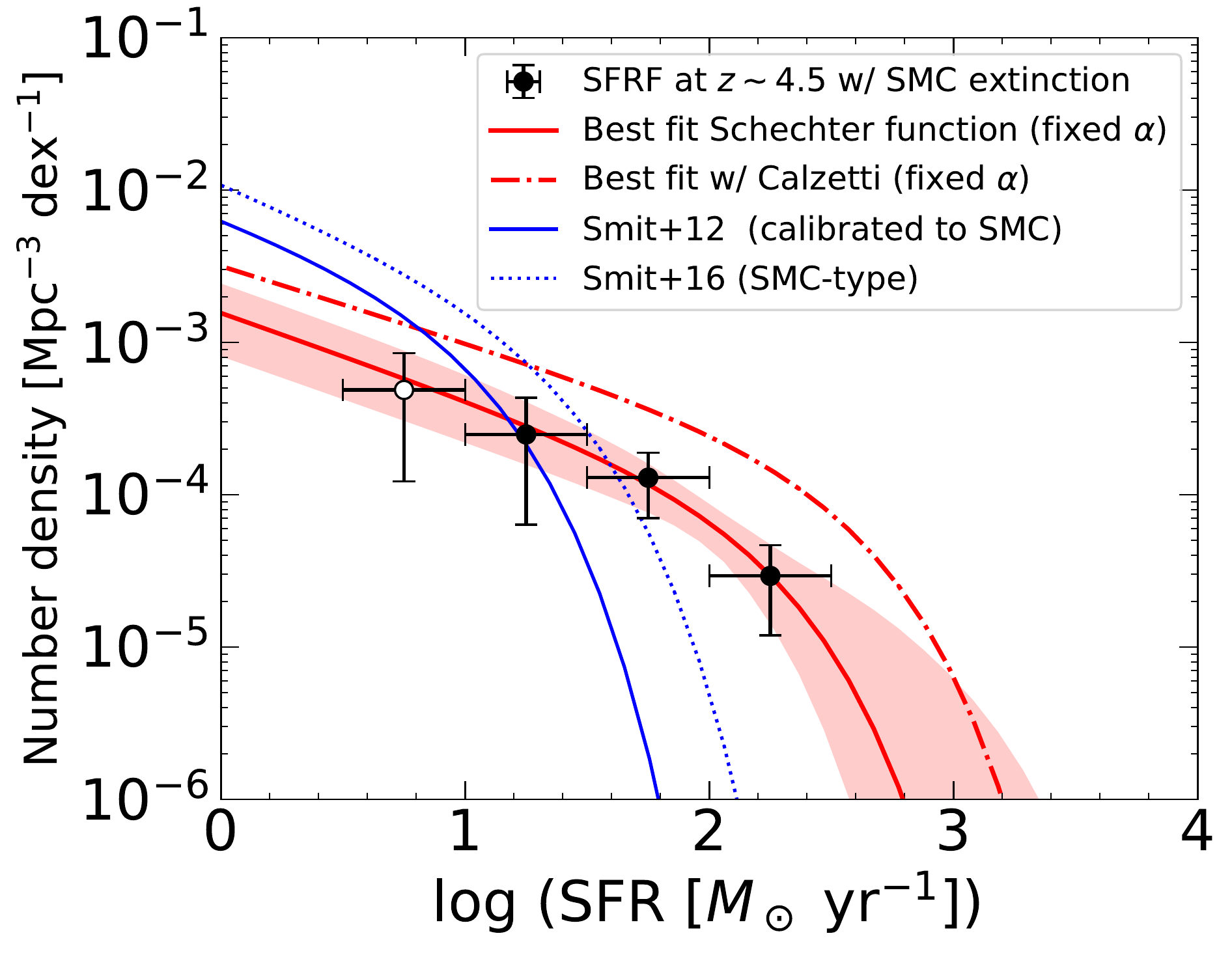}
\plotone{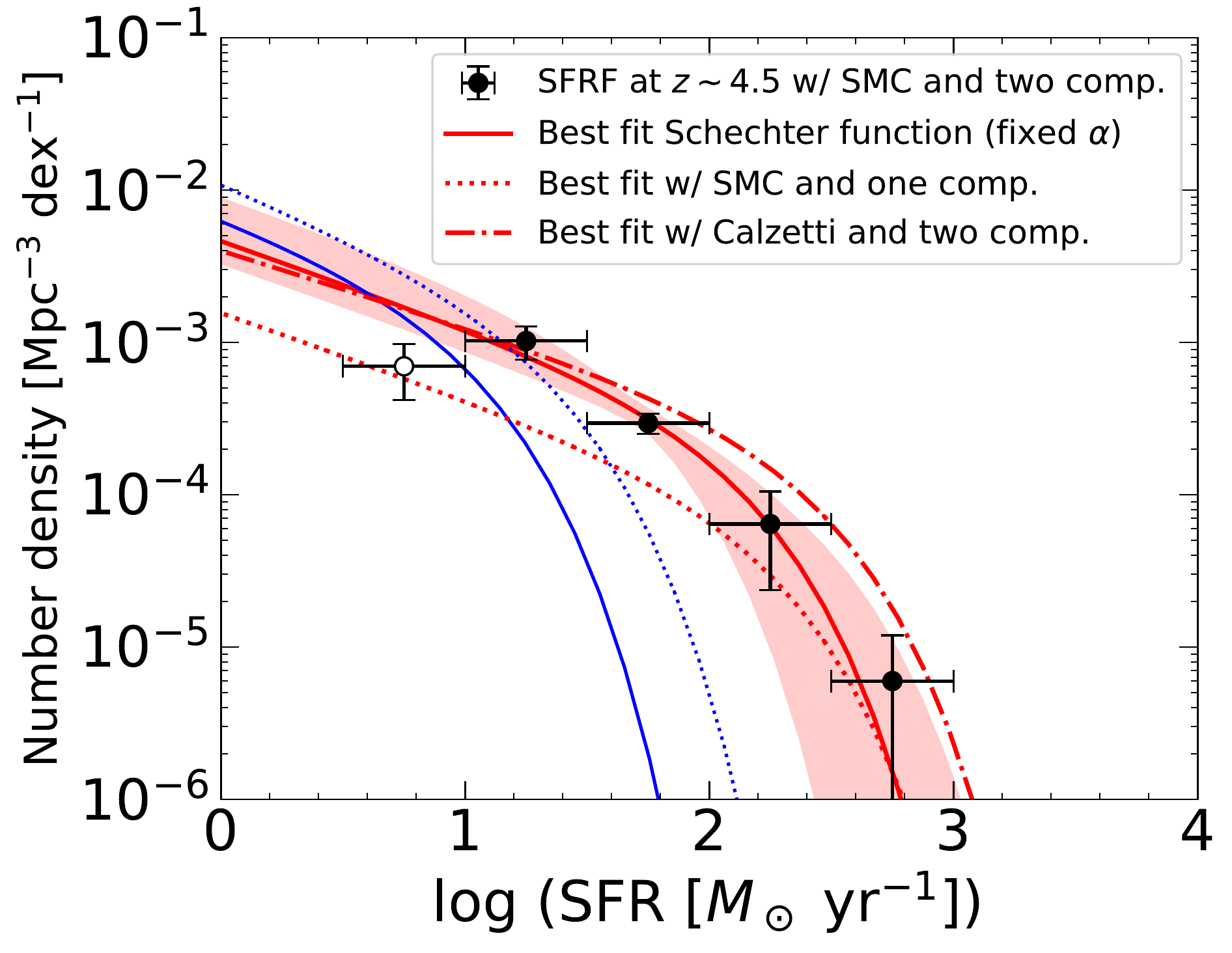}
\caption{SFRF derived with SMC extinction curve and the best-fit Schechter function (fixed $\alpha$, red solid line) for the one-component model (top) and the two-component model (bottom).
In both panels, red dash-dotted line shows the best-fit Schechter function to the SFRF with Calzetti extinction law with corresponding model.
Blue solid and dotted line shows the SFRFs converted from UV LFs with SMC-type dust correction (see text for details).
}
\label{SFRF_fitting_SMC}
\end{figure}

We show the resulting SFRF in the top panel of Figure \ref{SFRF_fitting_SMC}.
We also plot two SFRFs converted from UV LFs for comparison.
In S12, UV LF is converted into SFRF assuming the Meurer relation (equation (\ref{IRX_M99})).
Since \citet{meurer_dust_1999} dust correction is different from that with SMC law, we obtain a SFRF following the procedure described in S12 using SMC-type relation (\ref{IRX_M99_SMC}),
\begin{equation}\label{IRX_M99_SMC}
    A_{1600} = 2.45 + 1.1\beta
\end{equation}
instead of equation (\ref{IRX_M99}), and plot it in Figure \ref{SFRF_fitting_SMC}.
We also plot the SMC-type model by S16.

The completeness limit of our SFRF (SFR$\gtrsim10$ $M_\odot$ yr$^{-1}$) does not change with the extinction law.
Similar to the case of Calzetti attenuation, our result shows an excess to the SFRF estimated from UV LF.
We also fit Schechter function to our result fixing the faint-end slope to $\alpha = -1.56$, which is the same as that in S12 with SMC-type correction (equation (\ref{IRX_M99_SMC})).
We present the result of this fitting in the top panel of Figure \ref{SFRF_fitting_SMC} and Table \ref{Tab1}.

As we discussed in Section \ref{subsubsec:twocomp}, it is meaningful to explore if this excess can be reduced with two component model.
The result is shown in the bottom panel of Figure \ref{SFRF_fitting_SMC}.
Akin to the result in Section \ref{subsubsec:twocomp}, this excess is reduced only slightly and still exists even with the two component model.
In the case of SMC extinction law, the effect of increasing the value of SFR by adopting two-component model is distinctive, and the number density of galaxies with SFR $\sim10^{1-2}$ $M_\odot$ yr$^{-1}$ increases significantly.

\begin{figure}[tpb]
\plotone{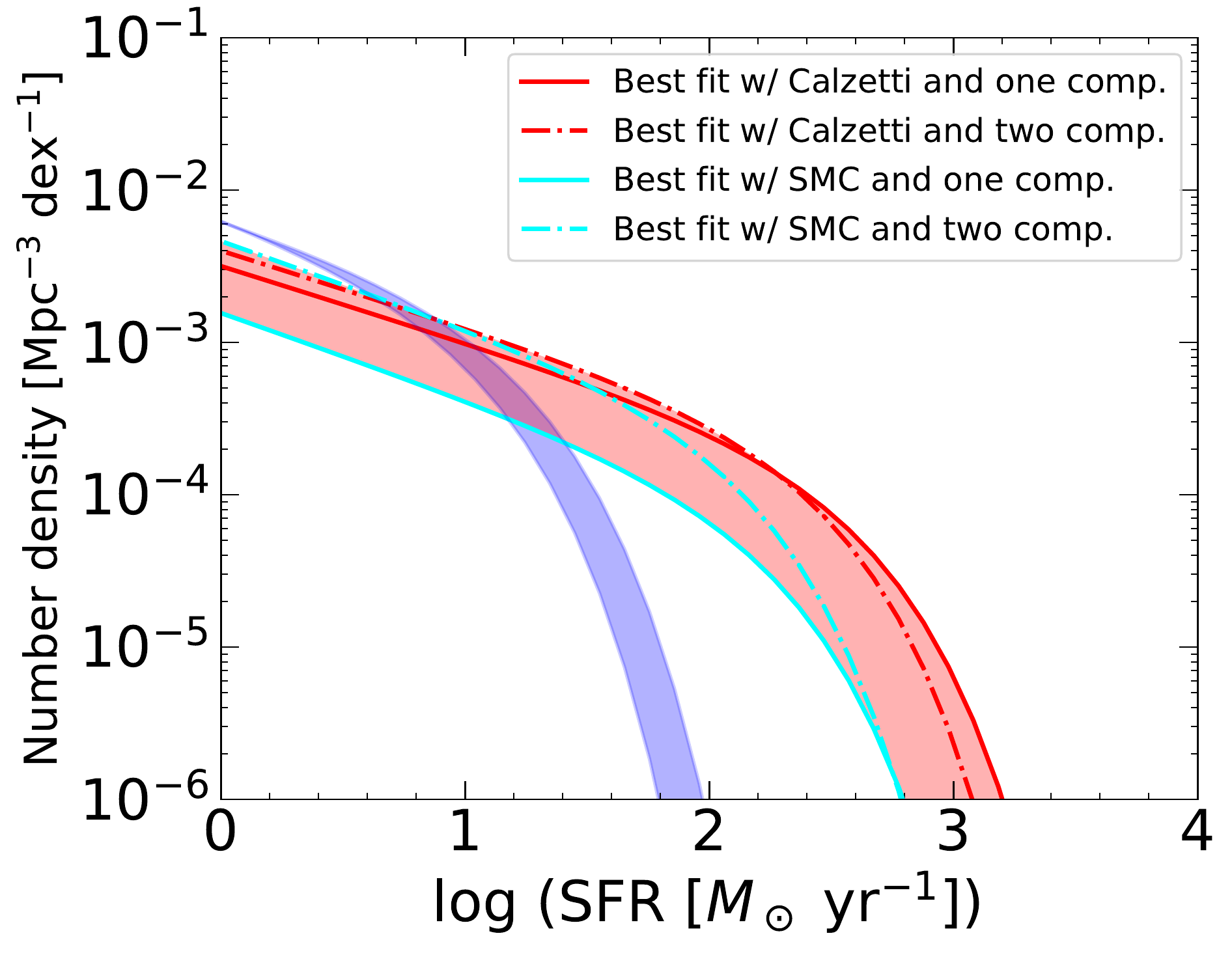}
\caption{All of the SFRFs derived in this work.
Red shaded region shows the area between the maximum and minimum of the SFRFs, where we expect the true SFRF should locate.
Blue shaded region shows the area between the SFRF that is estimated from UV LF with Calzetti extinction and that with SMC extinction (S12).
}
\label{SFRFs_all}
\end{figure}

We conclude that SFRF derived from SED fitting shows a significant excess to that estimated from UV LF in the large SFR ($\gtrsim10$ $M_\odot$ yr$^{-1}$), and this is robust regardless of the choice of dust extinction law or stellar composition models including various SFHs and two component model.
The SFRF can vary especially with the assumption of dust extinction curve, so we expect that the true value of SFRF locates between the maximum and minimum among the four SFRFs derived here.
In Figure \ref{SFRFs_all}, we show all of these SFRFs and the region where we expect the true value locates.

\subsection{Further inspections on the excess}\label{subsec:further_insp}
\subsubsection{Difference in UV luminosity function?}
The difference of SFRF may be originated in the difference of UV LF due to such as field-to-field variance.
We derive (dust-uncorrected) rest UV LF using our samples to see whether the UV LF is the same as those in the previous studies.
The rest-UV magnitude $M_{UV}$ of each galaxy is calculated using the apparent magnitude in $i_{814}$ band and its photometric redshift.
We derive UV LF with both of the target sample and final sample.
With final sample, the UV LF $\phi(M_{UV})$ [Mpc$^{-3}$ mag$^{-1}$] is estimated using the same equation as we used for SFRF (equation (\ref{estimate_density_per_dex})).
With the target sample, since this sample is not affected by the S/N cut and elimination of blended objects in the IRAC 4.5 $\mu$m band, the UV LF is estimated without corresponding corrections, $f^{\mathrm{sel}}$ and $f^{\mathrm{det}}(m_{ch2,i})$.

\begin{figure}[tpb]
\plotone{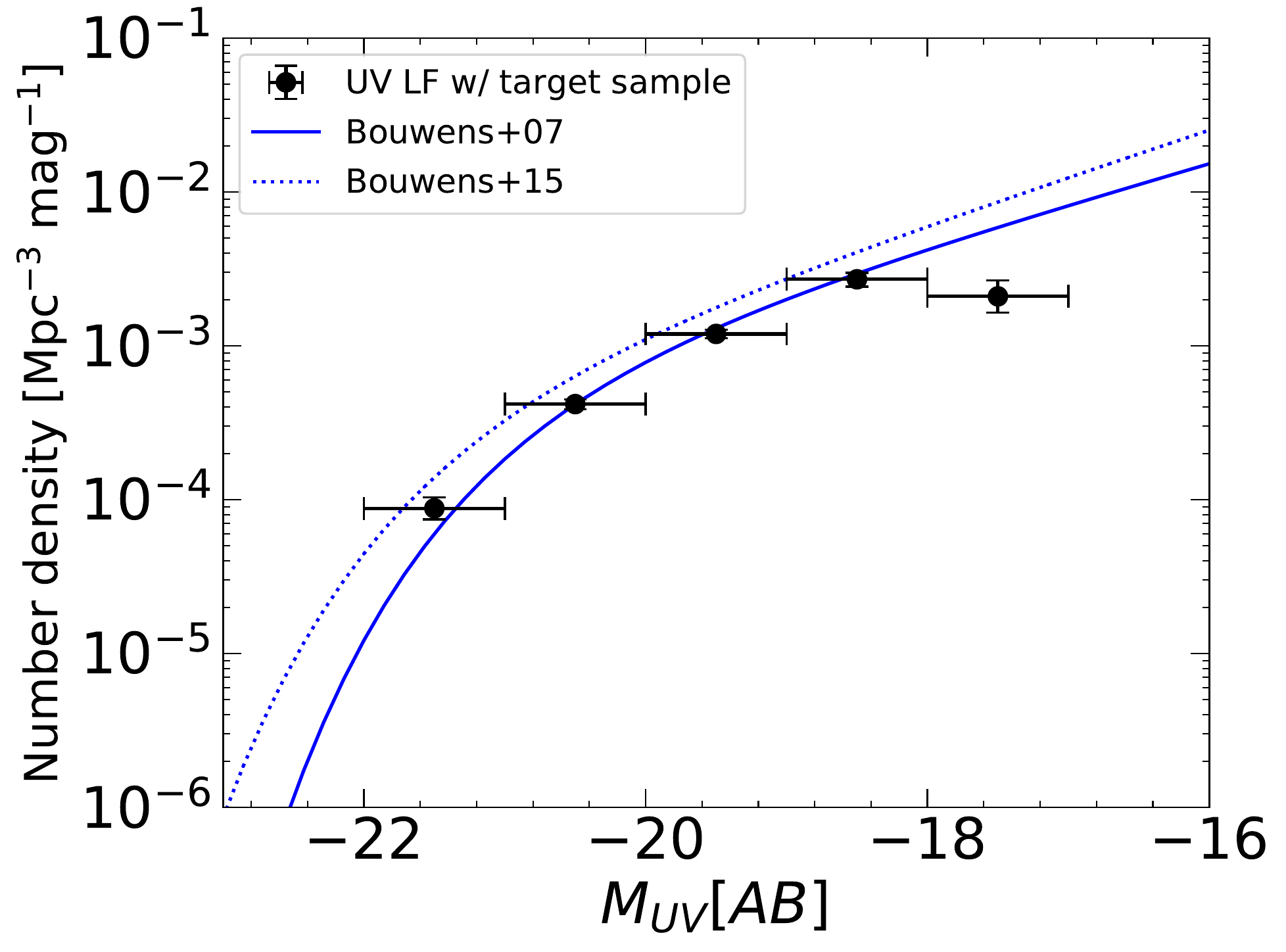}
\plotone{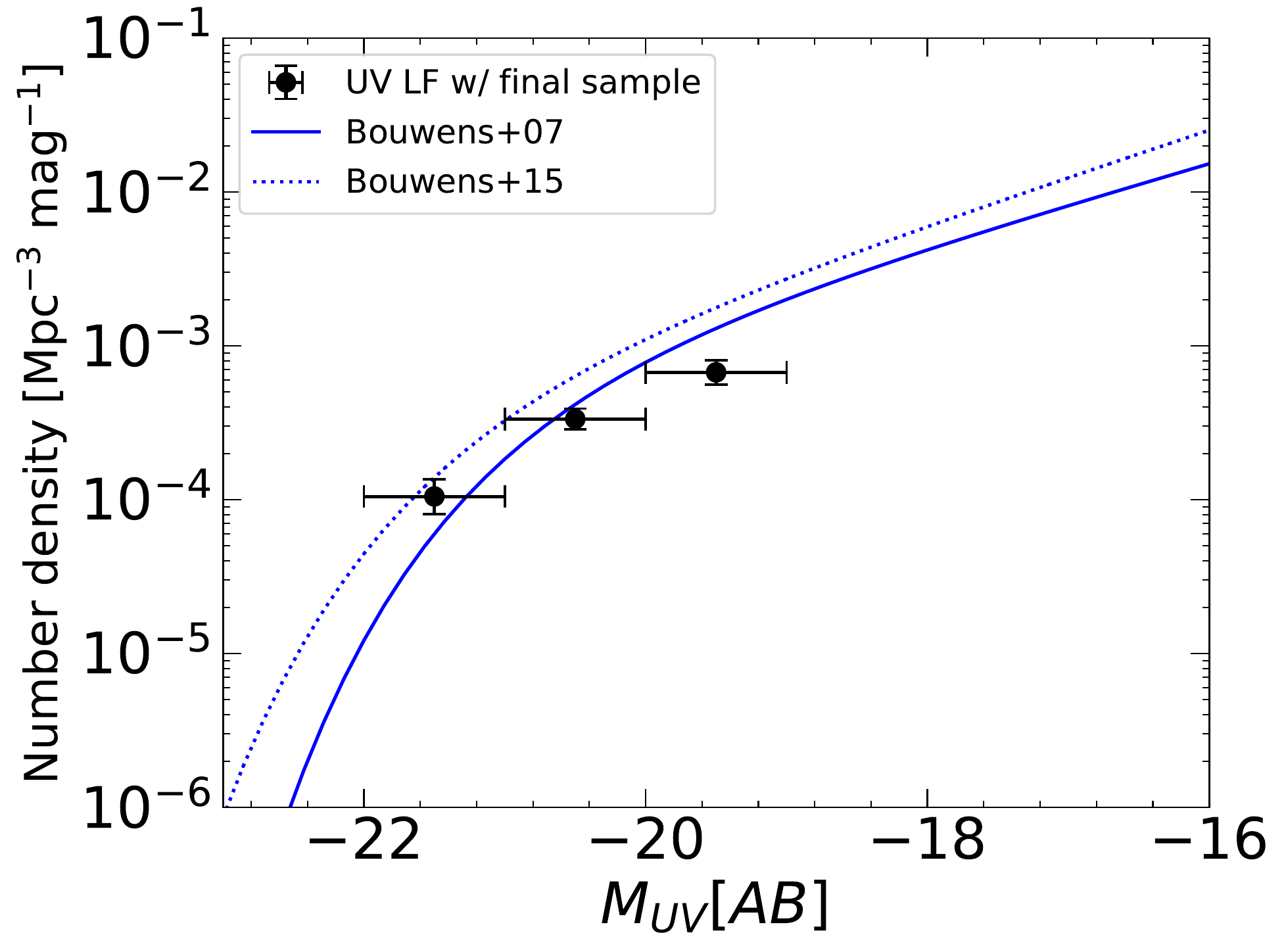}
\caption{Dust-uncorrected UV LFs with the target sample (top) and final sample (bottom).
The completeness limit of final sample, $\mathrm{SFR} >10$ $M_\odot$ yr$^{-1}$, corresponds to $M_{UV} \lesssim -20$.
Blue solid and dotted line shows UV LFs from \citet{bouwens_uv_2007} and \citet{bouwens_uv_2015}, respectively.
}
\label{UVLFs}
\end{figure}

We show the result in Figure \ref{UVLFs}.
Since the SFRFs in S12 and S16 are derived based on the (dust-uncorrected) UV LF given by \citet{bouwens_uv_2007} and \citet{bouwens_uv_2015}, respectively, we plot them for comparison.
We can see that the UV LFs derived here are in good agreement with those we used for comparisons.
Furthermore, if we correct for the dust extinction using equation (\ref{IRX_M99}) for each galaxy, the (dust-corrected) rest-UV LF is almost identical to that is obtained by S12.
These indicate that the excess of SFRF is not caused by the difference of UV LF.
Note that the UV LF with final sample is lower at $M_{UV}=-19.5$, but this is due to the incompleteness of our final sample in the faint region ($\mathrm{SFR} <10$ $M_\odot$ yr$^{-1}$).
In addition, UV LFs with the final sample and target sample agree well with each other in the bright region ($\mathrm{SFR} >10$ $M_\odot$ yr$^{-1}$), which suggests that the incompleteness due to the S/N cut and elimination of blended objects in the IRAC 4.5 $\mu$m band is well corrected with our method.

\subsubsection{Consistency with FIR observation}
It is important to see the consistency between our result and FIR observation.
About one third ($\sim 69$ arcmin$^2$) of CANDELS GOODS-South field is observed with ALMA band 6 in GOODS-ALMA project \citep{franco_goods-alma_2018,franco_goods-alma_2020}.
They presented a catalog of galaxies detected at 1.1 mm with a flux density limit of 640 $\mu$Jy ($\sim3.5 \sigma$).
At the redshift of $z\sim4.5$, this corresponds to a monochromatic luminosity limit ($\nu L_\nu$ unit) of $9.2\times 10^{10} L_\odot$ at $\lambda_{\mathrm{rest}}\sim200$ $\mu$m.
Using SED template presented by \citet{schreiber_dust_2018}, this limit is converted into a total IR luminosity limit, and then into a SFR limit with the following equation by \citet{madau_cosmic_2014}
\begin{equation}
    \mathrm{SFR} = \kappa L_{\mathrm{TIR}}
\end{equation}
where $\kappa = 1.08\times10^{-10}$ $M_\odot$ yr$^{-1}$ $L_\odot^{-1}$ for Chabrier IMF.
The resulting SFR limit is $\sim 270$ $M_\odot$ yr$^{-1}$.

Among galaxies in our final sample, 52 galaxies are located in the region covered by GOODS-ALMA observation.
No counterpart is found in this ALMA-selected catalog.
Considering the variety of SFRs with the assumption of models\footnote{We have 7 models for SFHs, 2 models for dust extinction and 2 models for stellar composition.
Thus, we derived 28 values of SFR per one galaxy.}, there is no galaxy whose SFR is definitely larger than $\sim 270$ $M_\odot$ yr$^{-1}$ among the 52 galaxies.
Therefore, the lack of the counterpart is consistent with the SFRs estimated in this study.

\section{Cosmic Star Formation Rate Density}\label{Sec:CSFRD}
We calculate CSFRD ($\rho_{\mathrm{SFR}}$) using our best-fit Schechter function.
This is often provided by integrating UV LF down to some lower bound, and $M_{\mathrm{UV}}=-17.0$ mag is one of the common values of this lower bound.
UV luminosity is converted into SFR using equation (\ref{Madau98_relation})
\citep[][]{madau_star_1998}:
\begin{eqnarray}\label{Madau98_relation}
    L_{\mathrm{UV}} = 8.0\times10^{27} \frac{\mathrm{SFR}}{M_\odot\ \mathrm{yr}^{-1}}\ \mathrm{erg}\ \mathrm{s}^{-1} \mathrm{Hz}^{-1}
\end{eqnarray}
Note that this value is for Salpeter IMF, so we divide it by a constant factor of 0.63 \citep{madau_cosmic_2014} to convert into Chabrier IMF.
We integrate our SFRF down to $\sim 0.22$ $M_\odot$ yr$^{-1}$, equivalent to $M_{\mathrm{UV}}=-17.0$ mag.

\begin{figure}[tpb]
\includegraphics[width=1.0\columnwidth, angle=0]{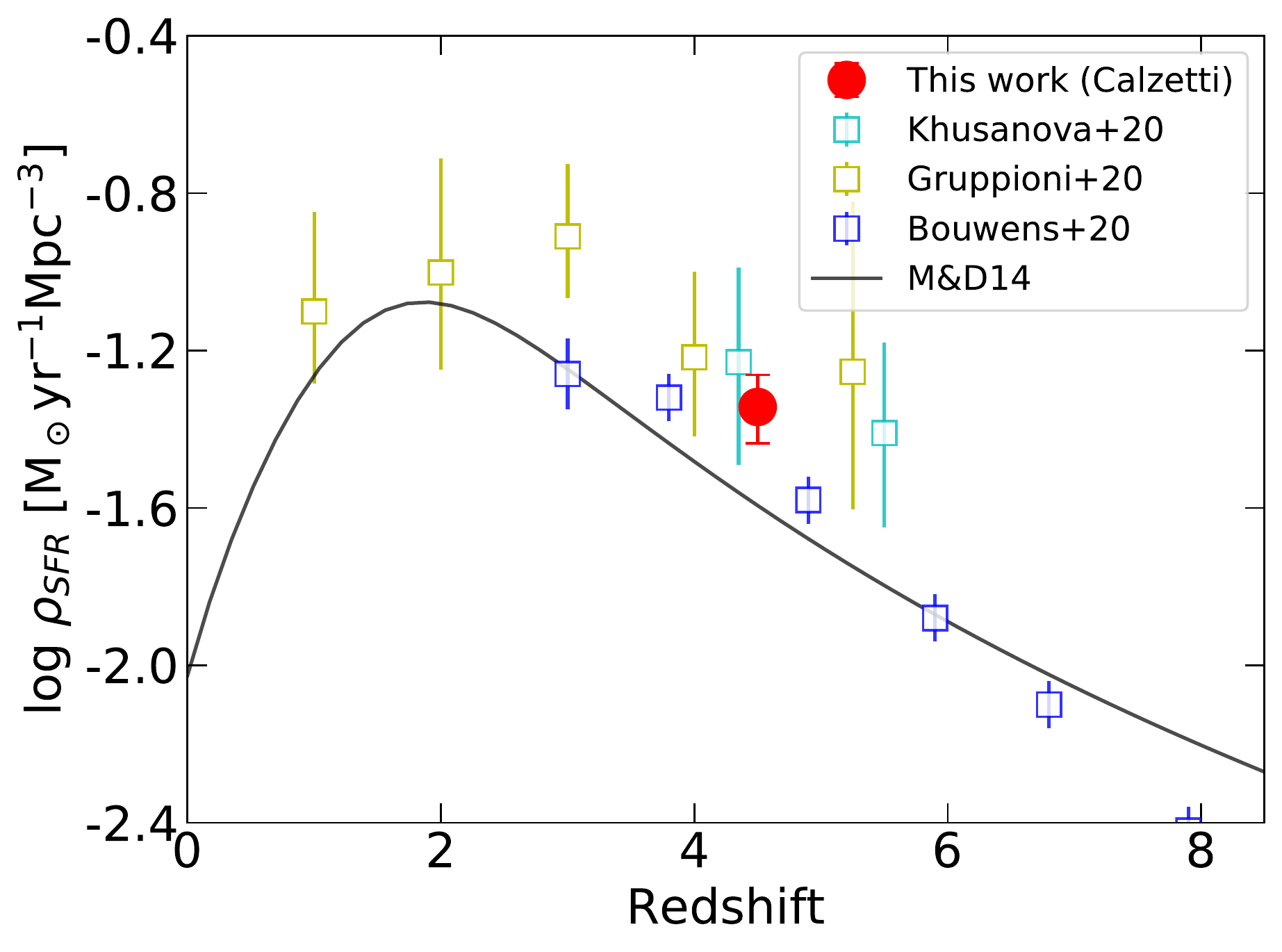}
\caption{Redshift evolution of CSFRD.
Our result is shown by the red filled circle.
Black solid line shows the best-fit for the redshift evolution of CSFRD by \citet{madau_cosmic_2014} converted into Chabrier IMF, and some recent studies from the literature are also presented by open square for comparison (\citealt{khusanova_alpine-alma_2020}, \citealt{gruppioni_alpine-alma_2020}, \citealt{bouwens_alma_2020}).
Results by \citet{khusanova_alpine-alma_2020} and \citet{gruppioni_alpine-alma_2020} are based on FIR observations, and that by \citet{bouwens_alma_2020} is based on UV observation.
For clarity, the result by \citet{khusanova_alpine-alma_2020} at $z=4.5$ is shifted $\Delta z = -0.15$.
}
\label{CSFRD}
\end{figure}

Resulting CSFRD is shown in Figure \ref{CSFRD} and Table \ref{Tab1}.
The uncertainty for the CSFRD is taken to be the maximum and minimum values of CSFRD with the parameters within the $1\sigma$ contour (e.g., Figure \ref{SFRF_fitting}).
In Figure \ref{CSFRD}, we plot the integrated value of the best-fit function for the SFRF with Calzetti law and the one-component model fixing the faint-end slope $\alpha$ as a fiducial one.
As can be seen in Figure \ref{CSFRD}, the excess of our SFRF makes CSFRD increase by $\sim0.25$ dex as compared with that estimated from UV LF at $z\sim4.5$ \citep{madau_cosmic_2014}.
Recent studies from FIR observations (\citealt{khusanova_alpine-alma_2020}, \citealt{gruppioni_alpine-alma_2020}) are also plotted in Figure \ref{CSFRD}, and our result is comparable to them.

Since our SFRF is derived in $\mathrm{SFR} >10$ $M_\odot$ yr$^{-1}$, we also calculate the CSFRD without extrapolating to $0.22$ $M_\odot$ yr$^{-1}$.
The CSFRDs integrated down to $10$ $M_\odot$ yr$^{-1}$ ($\rho^{\mathrm{complete}}_{\mathrm{SFR}}$) are shown in Table \ref{Tab2}.
These values are smaller than the CSFRDs obtained with the extrapolation by $\sim0.1$ dex, but still larger than CSFRD estimated from UV LF by $\sim0.2$ dex.
$\rho^{\mathrm{complete}}_{\mathrm{SFR}}$ occupies $\sim80\ \%$ of $\rho_{\mathrm{SFR}}$ (Table \ref{Tab2}), suggesting SF activity at this redshift is dominant by intensively star-forming galaxies, though the faint-end slope is not constrained well.

\begin{deluxetable}{cccc}
\tablenum{2}\label{Tab2}
\tablecaption{CSFRD of complete region }
\tablewidth{0pt}
\tablehead{
\colhead{Model} & \colhead{$\log_{10}\rho_{\mathrm{SFR}}$} & \colhead{$\log_{10}\rho^{\mathrm{complete}}_{\mathrm{SFR}}$} & \colhead{Fraction} \\
\colhead{} & \colhead{($M_\odot$ yr$^{-1}$ Mpc$^{-3}$)} & \colhead{($M_\odot$ yr$^{-1}$ Mpc$^{-3}$)} &
}
\decimalcolnumbers
\startdata
C,1,f & $-1.34^{+0.08}_{-0.09}$ & $-1.42^{+0.09}_{-0.11}$ & 0.84\\
C,1,v & $-1.36^{+0.19}_{-0.10}$ & $-1.42^{+0.11}_{-0.09}$ & 0.86\\
C,2,f & $-1.33^{+0.07}_{-0.09}$ & $-1.42^{+0.07}_{-0.09}$ & 0.80\\
S,1,f & $-1.91^{+0.13}_{-0.16}$ & $-2.05^{+0.19}_{-0.17}$ & 0.72\\
S,2,f & $-1.50^{+0.10}_{-0.09}$ & $-1.66^{+0.11}_{-0.13}$ & 0.69
\enddata
\tablecomments{The first column indicates the assumptions in deriving SFRF and fitting Schechter function. The first letter shows dust extinction (C$=$Calzetti, S$=$SMC), the second shows stellar component model (1$=$one component, 2$=$two component), and the third shows the treatment of faint-end slope (f$=$fixed, v$=$varied).
The second and third column indicates the CSFRD integrated down to $0.22$ and $10$ $M_\odot$ yr$^{-1}$, respectively, and the fourth column indicates the fraction of $\rho^{\mathrm{complete}}_{\mathrm{SFR}}$ to $\rho_{\mathrm{SFR}}$.}
\end{deluxetable}

In addition, to examine the effect of the choice of the dust extinction law on CSFRD ($\rho^{\mathrm{complete}}_{\mathrm{SFR}}$ or $\rho_{\mathrm{SFR}}$), we see the difference of CSFRD when the assumption of the dust extinction changes while the other assumptions such as stellar component model are fixed.
With one-component model, as can be seen in Table \ref{Tab2}, adopting SMC extinction law reduces $\rho^{\mathrm{complete}}_{\mathrm{SFR}}$ and $\rho_{\mathrm{SFR}}$ by $\sim0.6$ dex as compared with that obtained from Calzetti law (c.f. "C,1,f" and "S,1,f" model in Table \ref{Tab2}).
However, with two-component model, the difference of CSFRD between Calzetti and SMC extinction law is only $\sim0.2$ dex (c.f. "C,2,f" and "S,2,f" model in Table \ref{Tab2}).
Thus, the choice of dust extinction law seems to affect CSFRD, but the amount of it is still uncertain.
Note that these values of difference ($\sim0.6$ dex and $\sim0.2$ dex) do not change with the range of integration, i.e. $\rho^{\mathrm{complete}}_{\mathrm{SFR}}$ or $\rho_{\mathrm{SFR}}$, so this uncertainty is not due to the poor constraint of our SFRF on faint-end slope.

Next, we intend to see the effect of our photo-$z$ selection.
As we described in Section \ref{subsec:sample_sel}, we extracted objects that meet the criterion of equation (\ref{photo-z_criteria}).
Only about a half of objects whose photometric redshift $z_{best}$ is nominally in our target redshift range $3.88<z<4.94$ can pass this criterion (Figure \ref{sample_flow}).
Thus, our photo-$z$ selection may have non-negligible effect on our result.

To examine this, we set an alternative criterion as
\begin{equation}\label{Eqn:nominal_photz}
    3.88 < z_{best} < 4.94
\end{equation}
instead of equation (\ref{photo-z_criteria}) (we show this alternative sample selection by black dotted arrow in Figure \ref{sample_flow}) and derive SFRF with this alternative sample of galaxies following the same procedure as we did in Section \ref{subsec:SFRF}.
As a result, SFRF in the complete region ($\mathrm{SFR} >10$ $M_\odot$ yr$^{-1}$) is systematically larger than that shown in Figure \ref{SFRF_fitting} by $\sim0.3$ dex, i.e., $\phi^*$ and CSFRD is larger by $\sim0.3$ dex.
The inclusion of the objects which meet equation (\ref{Eqn:nominal_photz}) should contain galaxies not in our target redshift range.
Thus the effective volume should be larger than that we adopt.
However, it is difficult to asses the effect quantitatively.
Thus the value of $\sim0.3$ dex must be the upper limit for the impact of this effect.
It should be noted that the change of the SFRF here only affects on $\phi^*$, which results in the increase of $\rho_{\mathrm{SFR}}$ by the same amount.

\section{Summary}\label{Sec:Summary}
In deriving SFRF, correction for the dust extinction is a major concern.
Dust has much less effect on the rest optical light, so deriving SFRF with not only rest UV but also rest optical data can be an independent estimation of SFRF.
In this study, we derived a SFRF at $z\sim4.5$ down to $\sim10\ M_\odot\ \mathrm{yr}^{-1}$ based on photometric data from rest UV to optical of galaxies in the CANDELS GOODS-South field using SED fitting.
We extensively examine assumptions on SED fitting including various SFHs, dust extinction laws, and the two-component model.
Using the resulting SFRF, we also calculate the CSFRD by integrating to $0.22\ M_\odot\ \mathrm{yr}^{-1}$ which corresponds to $M_{\mathrm{UV}} = -17$ mag.
Our main results are as follows:
\begin{enumerate}
    \item The SFRF at $z\sim4.5$ derived from SED fitting shows an excess to that estimated from UV LF.
    As compared with the SFRF estimated from UV LF, the number density is larger by $\sim 1$ dex at a fixed SFR, or the best-fit Schechter parameter of $\mathrm{SFR}^*$ is larger by $\sim 1$ dex (Section \ref{subsec:SFRF}, Figure \ref{SFRF_fitting}).
    This result does not change with the choice of models such as various SFHs and one- or two-component model. (Section \ref{subsec:otherSFRF}, Figure \ref{SFRF_fitting_twocomp} and Table \ref{Tab1}).
    \item The SFRF varies with the choice of the dust extinction law, but the excess to UV-based SFRF still exists (Section \ref{subsec:otherSFRF}, Figure \ref{SFRF_fitting_SMC} and Table \ref{Tab1}).
    \item The CSFRD is calculated to be $4.53^{+0.94}_{-0.87}\times10^{-2}\ M_\odot\ \mathrm{yr}^{-1}\ \mathrm{Mpc}^{-3}$ with Calzetti extinction law.
    The excess of SFRF leads to an increase in the CSFRD by $\sim0.25$ dex as compared with that estimated from UV LF at $z\sim4.5$ (Section \ref{Sec:CSFRD} and Figure \ref{CSFRD}).
    However, the CSFRD varies with the choice of the dust extinction law (Section \ref{Sec:CSFRD} and Table \ref{Tab2}).
    \item The CSFRD is largely occupied ($\sim80\%$) by intensively star-forming ($\mathrm{SFR} > 10\ [M_\odot$ yr$^{-1}]$) galaxies regardless of the model assumption (Section \ref{Sec:CSFRD} and Table \ref{Tab2}).
\end{enumerate}

We thank the anonymous referee for his/her useful comments.
KO is supported by JSPS KAKENHI Grant Number JP19K03928.
FM is supported by a Research Fellowship for Young Scientists from the Japan Society of the Promotion of Science (JSPS).
This work is based on observations taken by the CANDELS Multi-Cycle Treasury Program with the NASA/ESA HST, which is operated by the Association of Universities for Research in Astronomy, Inc., under NASA contract NAS5-26555.

\appendix

\section{Two component model}\label{Appendix:Twocomp}
In this appendix, we describe the two-component model used in Section \ref{subsubsec:twocomp}.
This model consists of relatively old stellar population without star formation and young star-forming population.

As we discussed in Section \ref{subsec:reason}, the large discrepancy of SFRF is considered to be due to the large value of $E(B-V)$ derived from SED fitting.
Thus we introduce the two-component model to see whether the red color of such objects can be explained not by dust extinction but the continuum jump by old stars.
In the two-component SED fitting, it would be necessary to take an age of the old population and a fraction of the old population as additional free parameters.
However, we do not intend to examine the details of the two-component analysis and we just examine the rough behavior when we adopt the tow-component model.
Thus, we fix the age and fraction of the old population.

\begin{figure}[tpb]
\includegraphics[width=1.0\columnwidth, angle=0]{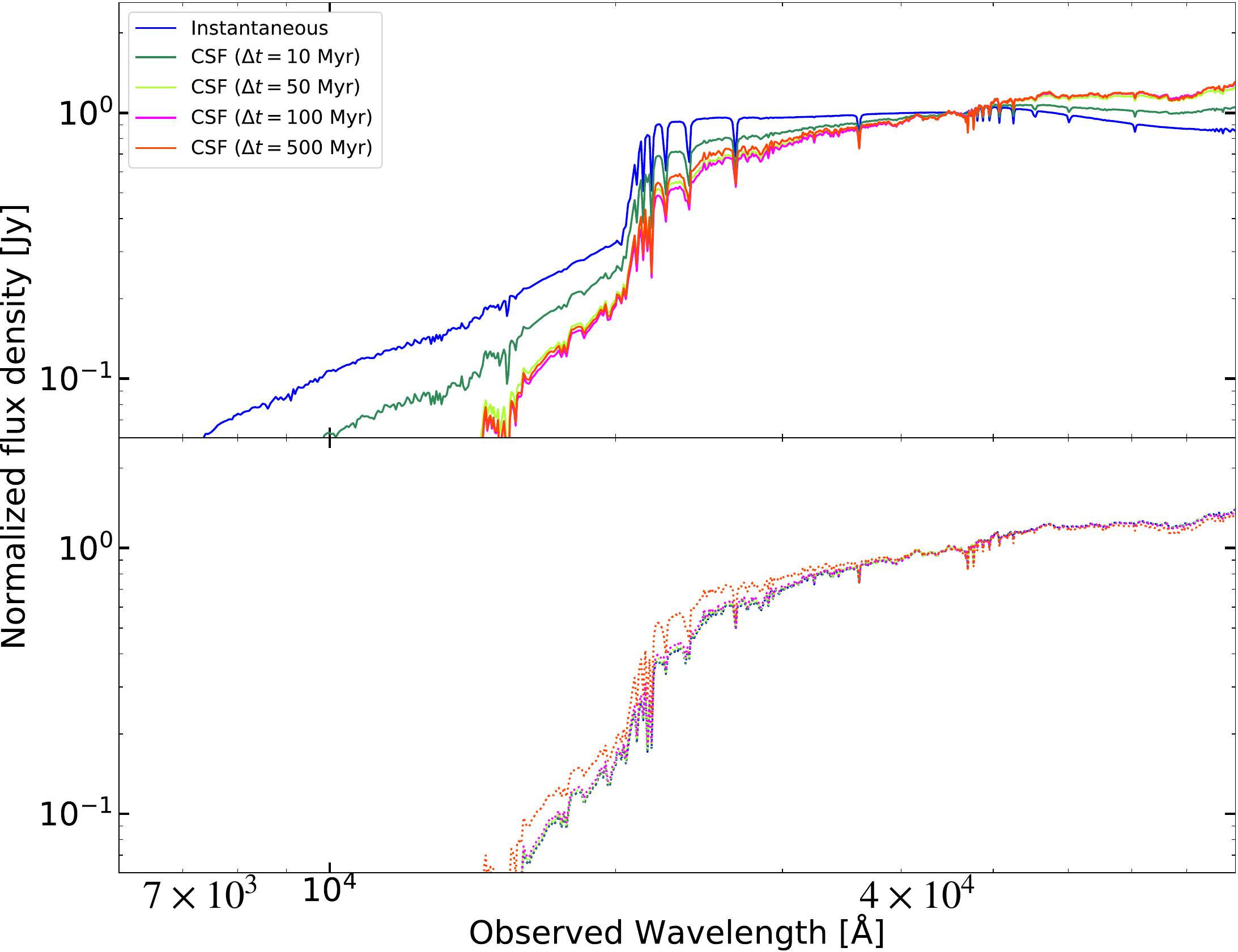}
\caption{Comparison of model spectra of old stellar component at $z=4.5$. All the spectra is normalized at 4.5 $\mu$m.
\textit{Top}: Model spectra with chemical evolution.
Blue, green, yellow, magenta and red line present the spectrum at the age of 1 Gyr of instantaneous strar formation, CSF with $\Delta t=10,\ 50,\ 100,\ 500$ Myr, respectively.
\textit{Bottom}: Model spectra without chemical evolution.
The metallicity is fixed to the Solar value.
Colors are the same as the top panel.
}
\label{Old_component_all}
\end{figure}

The age of the universe at $z\sim4.5$ is $\sim1.3$ Gyr, thus the age of old component is less than 1.3 Gyr.
We then see which SFH makes their continuum reddest with 1 Gyr old\footnote{The onset of star formation of 1 Gyr old stellar population at $z\sim4.5$ is $z\sim13$.}.
We compare stellar continuum spectra at the age of 1 Gyr with five different SFHs: instantaneous burst and CSF with the duration of $\Delta t =10,\ 50,\ 100,\ 500$ Myr from the onset of star formation using the population synthesis code \textsc{P\'egase}.3.
We show these spectra in Figure \ref{Old_component_all} (top panel).

Naively, instantaneous star formation history is expected to be the reddest.
As shown in Figure \ref{Old_component_all} (bottom panel), this is the case when the stellar metallicity is fixed.
However, the spectrum of CSF with the duration of $\Delta t=100$ or $500$ Myr is the reddest when we consider the chemical evolution.
This is essentially because of their stellar metallicity.
In \textsc{P\'egase}.3, the model SED is calculated following the chemical evolution, so the stellar population with instantaneous star formation history has extremely low (almost zero) metallicity and the SED shows a very blue color.
This effect is still seen for the CSF with $\Delta t=10$ Myr, and the SED becomes the reddest when the star formation continues for a reasonable duration.
Therefore, we adopt the spectrum of CSF model with $\Delta t=500$ Myr aged 1 Gyr as the old population spectrum.

Next, we examine the fraction of the old component to the whole system at 4.5 $\mu$m.
We test the fraction of $25\%, 50\%$ and $75\%$, and find that the fraction of $25\%$ is too small to affect the result, and $75\%$ shows larger effect to reduce the best-fit values of $E(B-V)$ than $50\%$.
Hence, we fix the fraction to $75\%$\footnote{If we take a much larger fraction, the SED of young population shows unrealistic shape.}.
We then subtract the old population from the observed SED and make the same SED fitting to the residual as we did in Section \ref{Sec:SED_fitting}.
Here, we assume that there is no dust extinction in the old component.

\begin{figure*}[tpb]
\plotone{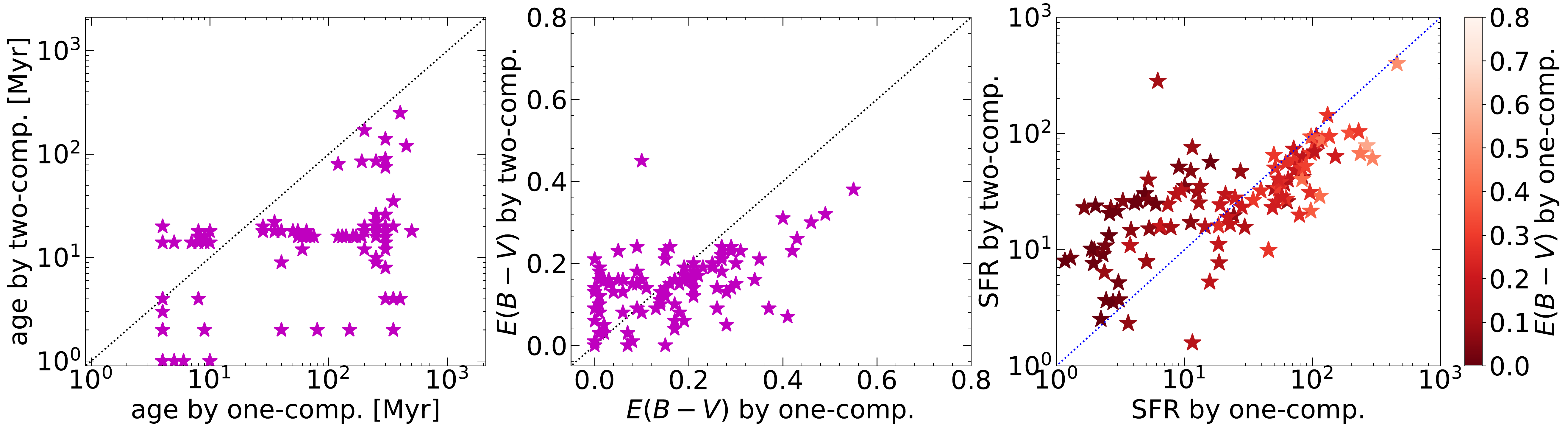}
\caption{Comparisons of age, $E(B-V)$ and SFR derived from SED fitting with the two-component model and one-component model. Dust extinction for young population is modeled with Calzetti law.
Only the case of exponentially declining SFH model with $\tau=100$ Myr is shown here, but the distribution does not change so much with SFHs.
\textit{Left}: Best-fit ages for the galaxies in our final sample. The age in the two-component model does not means the age of the galaxy but the age of young, star-forming population.
\textit{Center}: Best-fit $E(B-V)$. The $E(B-V)$ in the two-component model is for the young population.
\textit{Right}: SFR from SED fitting.
Symbols in this panel are colored according to the value of $E(B-V)$ derived with one-component model.
}
\label{2comp_comparison}
\end{figure*}
\begin{figure*}[tpb]
\plotone{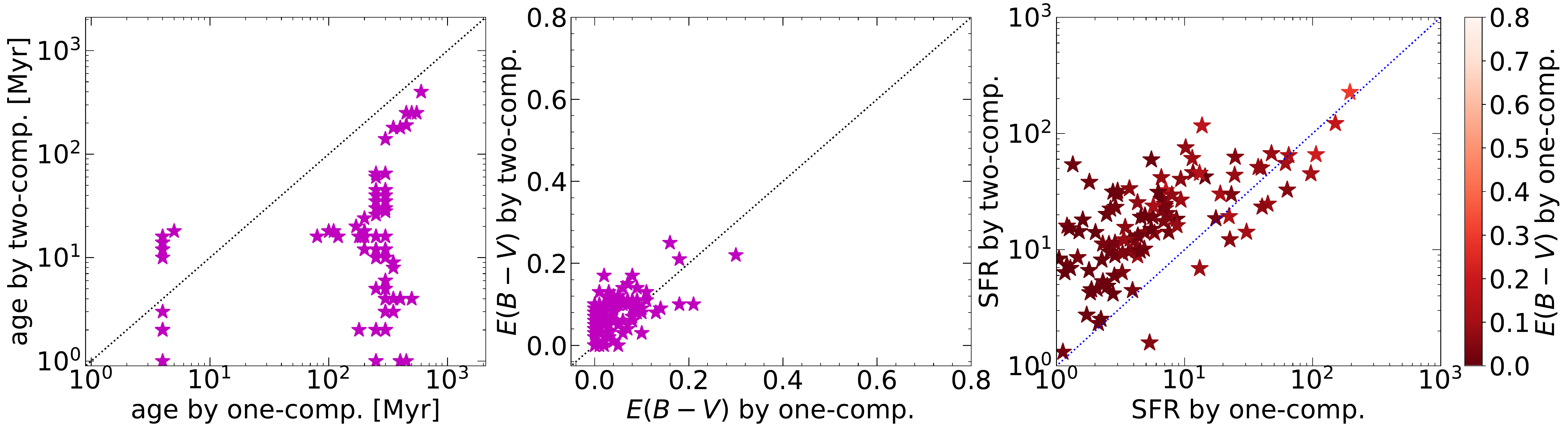}
\caption{Same as Figure \ref{2comp_comparison} but for SMC extinction curve for young population.
}
\label{2comp_comparison_SMC}
\end{figure*}
\begin{figure*}[tpb]
\plotone{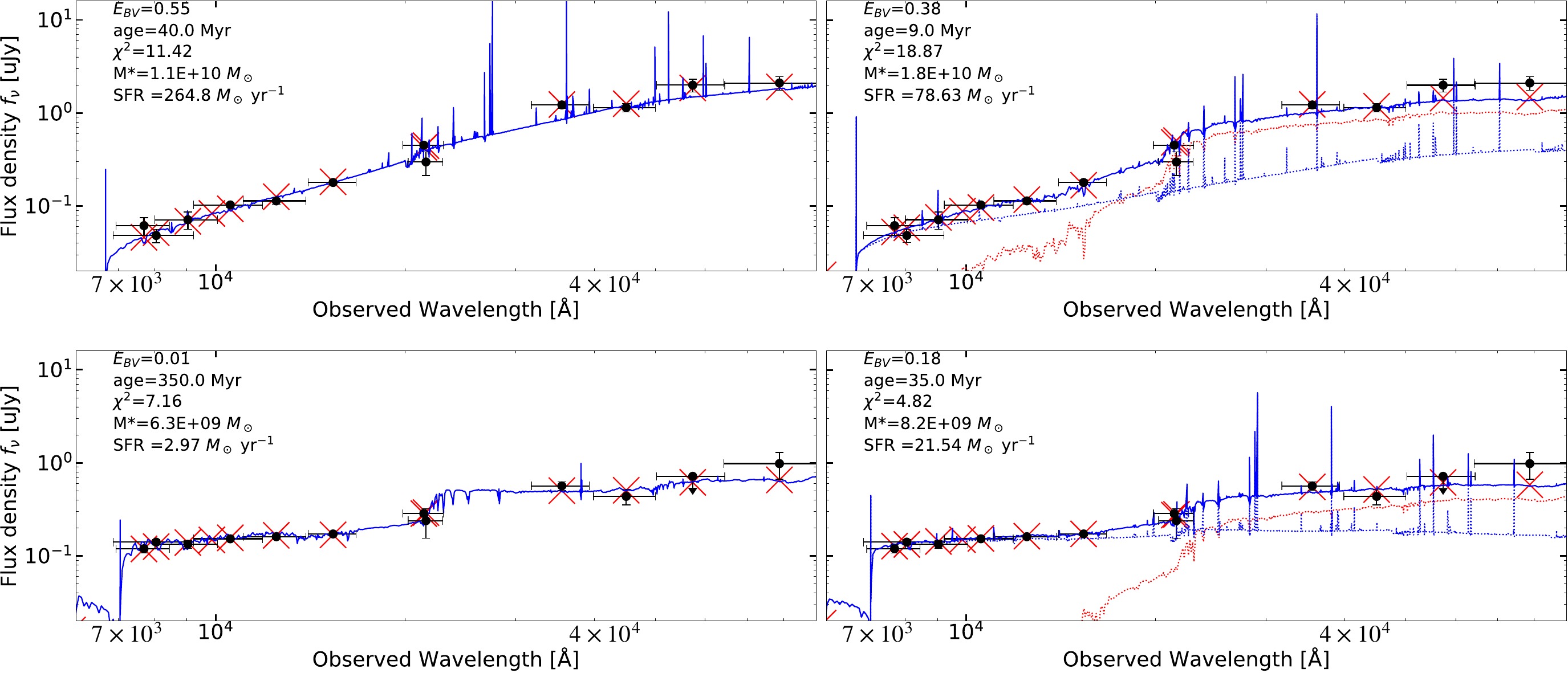}
\caption{Examples of the SED fitting.
\textit{Top}: An example galaxy whose best-fit $E(B-V)$ decreases in the two-component model.
Left and right panel show the best fit SED by one-component and two-component model, respectively.
\textit{Bottom}: An example galaxy whose best-fit $E(B-V)$ increases.
In both panels, black points with error bars represent the observed SED of each object.
Blue solid line represents the best-fit model spectrum of the entire galaxy, whose flux densities in each band are shown with red cross.
Blue and red dotted line in the right panel shows the best-fit spectra of young and old population, respectively.
We only show the result in the case of exponentially declining SFH model with $\tau=100$ Myr and Calzetti extinction law for one-component model and young population in the two-component model.
}
\label{SEDfit_ex}
\end{figure*}

The results are shown in Figure \ref{2comp_comparison} and \ref{2comp_comparison_SMC}. 
Note that the value of age and $E(B-V)$ in the two-component model is for the young population, not for the entire galaxy.
As expected, galaxies that have a large value of $E(B-V)$ in one-component model tend to show a lower value in the two-component model, and show a lower SFR by $\sim0-0.2$ dex.
However the difference is not so large, the two-component model can decrease the excess of SFRF only subtle.
We present an example of this kind of galaxy in the top panel of Figure \ref{SEDfit_ex}.



Interestingly, some galaxies show a larger $E(B-V)$ and SFR in the two-component model than in the one-component model.
An example of such galaxy is shown in bottom panel of Figure \ref{SEDfit_ex}.
We find that this kind of galaxy is, although there seems to be some excess in 3.6 $\mu$m band, interpreted as a almost passive galaxy in the one-component model because of its moderately red color from rest UV to optical.
In the two-component model, most of its rest optical light comes from the old population, thus the young population needs to be faint in rest optical, which prefers younger and bluer star-forming population.
On the other hand, the young population must reproduce the moderately red color of the rest UV, thus the young population has a moderate $E(B-V)$ and SFR gets larger than that in the one-component model.

In conclusion, when the two-component model is adopted, there seems to be two groups of galaxies;
in one group the best-fit values of $E(B-V)$ and SFR decrease, while in the other group the best-fit values of them increase.
Galaxies with SFR $\geq 10^{1.5}$ $M_\odot$ yr$^{-1}$ by the one-component model tend to belong to the former, and the two-component model reduces their best-fit SFR values by $\sim0-0.2$ dex.
Galaxies with SFR $\leq 10^{1.5}$ $M_\odot$ yr$^{-1}$ by the one-component model tend to belong to the latter, and the two-component model increases their SFR to $\sim 10^1$-$10^2$ $M_\odot$ yr$^{-1}$.
Therefore, adopting the two-component model reduces the number density of galaxies with the largest SFR ($\geq 10^2$ $M_\odot$ yr$^{-1}$) and increases that of galaxies with intermediate SFR ($\sim 10^1$-$10^2$ $M_\odot$ yr$^{-1}$) (c.f. Figure \ref{SFRF_fitting_twocomp} and \ref{SFRF_fitting_SMC}).

To determine whether such two-component model is the case or not, it is necessary to observe their rest optical light with a higher angular resolution and compare the spatial distribution between rest UV and optical, which is expected to correspond to star-forming component and old component, respectively.
\textit{JWST} will enable us to conduct such observations.

\clearpage
\newpage


\bibliography{my_library}{}
\bibliographystyle{aasjournal}



\end{document}